\documentclass[twocolumn]{aastex63}

\usepackage{amsmath}
\usepackage{graphicx}
\usepackage[caption=false]{subfig}

\begin{document}

\title{Prospects for Characterizing the Haziest Sub-Neptune Exoplanets with High Resolution Spectroscopy}

%% \author[xxxx-xxxx-xxxx-xxxx]{Author Name}
%%
%% This will hyperlink the author name to the author's ORCID page. Note that
%% during compilation, LaTeX will do some limited checking of the format of
%% the ID to make sure it is valid. If the "orcid-ID.png" image file is 
%% present or in the LaTeX pathway, the OrcID icon will appear next to
%% the authors name.
%%Please use multiple \affiliation calls for to document more than one affiliation.
%%
%% The new \altaffiliation can be used to indicate some secondary information
%% such as fellowships. This command produces a non-numeric footnote that is
%% set away from the numeric \affiliation footnotes.  
%% Use \email to set provide email addresses. Each \email will appear on its own line so you can put multiple email address in one \email call. A new \correspondingauthor command is available in V6.3 to identify the corresponding author of the manuscript. 
%%
%% If done correctly the peer review system will be able to
%% automatically put the author and affiliation information from the manuscript
%% and save the corresponding author the trouble of entering it by hand.

\correspondingauthor{Callie E. Hood}
\email{cehood@ucsc.edu}

\author[0000-0003-1150-7889]{Callie E. Hood}
\affiliation{Department of Astronomy \& Astrophysics, University of California,  Santa Cruz, CA 95064, USA}

\author[0000-0002-9843-4354]{Jonathan J. Fortney}
\affiliation{Department of Astronomy \& Astrophysics, University of California,  Santa Cruz, CA 95064, USA}

\author[0000-0002-2338-476X]{Michael R. Line}
\affiliation{School of Earth and Space Exploration, Arizona State University, Tempe, AZ 85287, USA}

\author[0000-0002-0618-5128]{Emily C. Martin}
\affiliation{Department of Astronomy \& Astrophysics, University of California,  Santa Cruz, CA 95064, USA}

\author[0000-0002-4404-0456]{Caroline V. Morley}
\affiliation{Department of Astronomy, University of Texas, Austin, TX 78712, USA}

\author[0000-0002-4125-0140]{Jayne L. Birkby}
\affiliation{Anton Pannekoek Institute of Astronomy, University of Amsterdam, Science Park 904, Amsterdam,1098 XH, The Netherlands}

\author[0000-0003-4408-0463]{Zafar Rustamkulov}
\affiliation{Department of Astronomy \& Astrophysics, University of California,  Santa Cruz, CA 95064, USA}
\affiliation{Department of Earth and Planetary Sciences, Johns Hopkins University, Baltimore, MD, USA}

\author[0000-0003-3444-5908]{Roxana E. Lupu}
\affiliation{BAER Institute / NASA Ames Research Center, Moffett Field, CA 94035, USA}

\author{Richard S. Freedman}
\affiliation{SETI Institute, Mountain View, CA 94043, USA}
\affiliation{NASA Ames Research Center, Moffett Field, CA 94035, USA}

\begin{abstract}

Observations to characterize planets larger than Earth but smaller than Neptune have led to largely inconclusive interpretations at low spectral resolution due to hazes or clouds that obscure molecular features in their spectra. However, here we show that high-resolution spectroscopy (R $\sim$ 25,000 to 100,000) enables one to probe the regions in these atmospheres above the clouds where the cores of the strongest spectral lines are formed. We present models of transmission spectra for a suite of GJ1214b-like planets with thick photochemical hazes covering 1 - 5 $\mu$m at a range of resolutions relevant to current and future ground-based spectrographs. Furthermore, we compare the utility of the cross-correlation function that is typically used with a more formal likelihood-based approach, finding that only the likelihood based method is sensitive to the presence of haze opacity. We calculate the signal-to-noise of these spectra, including telluric contamination, required to robustly detect a host of molecules such as CO, CO$_{2}$, H$_{2}$O, and CH$_{4}$, and photochemical products like HCN, as a function of wavelength range and spectral resolution. Spectra in \textit{M} band require the lowest S/N$_{res}$ to detect multiple molecules simultaneously. CH$_{4}$ is only observable for the coolest models ($T_{\rm{eff}} =$ 412 K) and only in the \textit{L} band.  We quantitatively assess how these requirements compare to what is achievable with current and future instruments, demonstrating that characterization of small cool worlds with ground-based high resolution spectroscopy is well within reach.

\end{abstract}



\section{Introduction} \label{sec:intro}
\indent NASA's \emph{Kepler} mission has discovered thousands of exoplanet candidates with sizes between that of Earth and Neptune \citep{Borucki2011,Batalha2013, Burke2014, Mullally2015, Rowe2015}. These sub-Neptune planets appear to be common, around both M dwarf and Sun-like stars \citep{Petigura2013, Fressin2013, Burke2015, Dressing2015}. In fact, around a third of Sun-like stars host a planet of this size with orbital periods less than 100 days \citep{Petigura2013, Fressin2013, Burke2015}. 

\indent The measured bulk density of these planets could be consistent with a range of compositions \citep{Figueira2009, Rogers2010, Nettelmann2011}. Correspondingly, sub-Neptunes may have a diversity of compositions from rocky to gas-rich as expected from formation and evolution modelling \citep{Fortney2013, Moses2013, Lopez2014}.  A wide range of atmospheres is expected from their bulk compositions \citep{Morley2017, Kempton2018}.  The history of how a planet accreted or outgassed its atmosphere, and its subsequent evolution, may be encoded in the abundances or ratios of molecular species in its atmosphere \citep{Oberg2011, Booth2017, Espinoza2017}.  Thus, constraining the atmospheric makeup of a sample of sub-Neptune planets may be the best way to understand how and out of what material these objects form. Due to their abundance, more sub-Neptune planets are likely to be found by the TESS mission around nearby, bright M-dwarfs \citep{Ricker2015,Sullivan2015, Barclay2018}, providing prime targets for atmospheric characterization with JWST, ARIEL, and large ground-based telescopes \citep{Louie2018, Kempton2018, Zellem2019}.

\indent However, these planets have proven hard to characterize; most atmospheric features that are detected are weaker than expected for a solar-metallicity, cloud-free atmosphere \citep{Fraine2014, Fu2017, Wakeford2017, Wakeford2019}. Some measurements have been unable to detect atmospheric features at all \citep[e.g.][]{Knutson2014a, Kreidberg2014}. Incorporating potentially muted features into yield calculations, \cite{Crossfield2017} find that the expected yield of \emph{TESS} planets amenable to characterization with \emph{JWST} is up to 7x worse than when assuming cloud-free conditions. 

\indent GJ 1214b, a $6.16 \pm 0.91$ M$_{\oplus}$ and $2.71 \pm 0.24$ R$_{\oplus}$ planet around a M4.5 star \citep{Charbonneau2009}, is the prototype of this planetary class, and most dramatic example a ``difficult'' atmosphere. Observations of GJ 1214b taken with ground-based instruments and \emph{HST} are consistent with a flat transmission spectrum \citep[e.g.][]{Bean2010,Bean2011,Kreidberg2014}. While early observations were inconclusive, \cite{Kreidberg2014} achieved the signal-to-noise necessary to rule out just a clear but high mean molecular weight atmosphere as the source of the flat transmission spectrum. Instead, significant gray aerosol opacity has been invoked as the source of muted features in transmission spectra, including that of GJ 1214b  \citep[e.g.][]{Crossfield2013,Kreidberg2014, Knutson2014a, Knutson2014b, Iyer2016, Sing2016}.

\indent Aerosols can absorb and scatter light \citep{Heng2013}, providing an extra opacity source that dampens absorption features in transmission spectra \citep[e.g.][]{Deming2013, Kreidberg2014, Kreidberg2018, Stevenson2016}. \cite{Morley2013} explored two types of aerosols expected to form in GJ 1214b's atmosphere---clouds from equilibrium chemistry and a photochemical haze layer from the destruction of CH$_{4}$---finding that either aerosol over a range of parameters could flatten the planet’s transmission spectrum. Further cloud formation work found that KCl and ZnS clouds can only be consistent with observations at high metallicities (1000x solar) and with strong atmospheric mixing (Kzz = 10$^{10}$ cm$^{2}$ s$^{-1}$) \citep{Morley2015, Gao2018}. In contrast, photochemical hazes could explain the observed HST observations with lower metallicities $\sim$ 50$\times$ solar \citep{Morley2015}. If an aerosol is the cause of the observed flat transmission spectra, JWST may allow us to characterize sub-Neptunes with it’s longer wavelength coverage and higher resolution than HST \citep{Greene2016, Mai2019}. 

\indent However, another potential avenue for studying these atmospheres is ground-based, high-resolution spectroscopy. Over the past decade, spectroscopy with a resolving power $\geq$ 25,000 has been used to characterize the composition, dynamics, and thermal structure of exoplanet atmospheres (see \citealt{Birkby2018} for a recent review of the technique and resulting detections). The large variations in the radial velocity of a close-in exoplanet relative to the host star allow for the Doppler-shifted planet spectrum to be disentangled from the relatively static lines of the host star's spectrum as well as from the spectral absorption features of Earth's atmosphere. At high spectral resolution, molecular band heads are resolved into unique groups of individual lines allowing for robust detections from matching these lines to theoretical models. Though first suggested by \cite{Deming2000}, \cite{Brown2001}, and \cite{Sparks2002}, \cite{Snellen2010} was the first robust detection of a molecule (CO) in a planet's atmosphere using this technique with CRIRES at the VLT. Since then, molecules such as CO and H$_2$O have been routinely detected for a variety of planets in emission and transmission \citep[e.g.][]{Rodler2012, Birkby2013, Brogi2014}. While most of these studies have been of hot Jupiters, a few focused on smaller planets have yielded upper limits on molecular abundances \citep{Crossfield2011, Esteves2017, Deibert2019}. 

\indent At high spectral resolution, the cores of the strongest molecular lines are formed very high up in the planet's atmosphere, possibly above whatever cloud or haze deck may obscure features at low-resolution \citep{deKok2014, Kempton2014}. Thus, planets whose atmospheres are completely obscured in a low-resolution transmission spectrum may still be successfully characterized at high-resolution \citep{Birkby2018}. \cite{Pino2018b} showed that not only is H$_{2}$O detectable in the presence of an aerosol for a typical hot Jupiter, but the relative cross-correlation strength across multiple wavelength ranges could be used to detect the aerosol's presence. 

\indent The aim of this paper is to quantitatively study the feasibility of detecting molecular features of the haziest sub-Neptune planets with high-resolution transmission spectroscopy. Is this achievable?  And specifically, is this a science case for current instruments, or only for instruments on upcoming Extremely Large Telescopes? We investigate how a range of observational parameters including signal-to-noise ratio, spectral resolution, and wavelength coverage affect the detection of various molecules. 

\indent Furthermore, as high-resolution spectroscopy has become more common, the best way to robustly report detection significances of molecules has been explored. \cite{Brogi2019} proposed a new log likelihood function to use when comparing observed spectra to a model in the place of the traditional cross-correlation function. We will compare the utility of these two metrics, motivating our choice of the log likelihood function for the majority of this work. However, this method may be affected by any ``missing'' molecules present in the observed atmosphere but absent in the model spectrum. For illustration, we will consider how HCN, a high abundance photochemical product,  affects the observed spectra and reported detection significances. 

\indent This work is structured as follows. In section 2, we describe how we generate transmission spectra for hazy GJ 1214b analogs and discuss how we quantify the significance of molecular detection with this technique. In Section 3, we give an overview of the prospects for observing CO, CO$_{2}$, H$_{2}$O, and CH$_{4}$ across a range of planetary insolation levels as well as the spectral resolution and wavelength coverage of the data.   A discussion of our results are presented in Section 4 and our conclusions are summarized in Section 5.

\section{Methods} \label{sec:methods}
\subsection{Model Atmosphere and Spectra} \label{subsec:trspec}
\begin{figure*}
\centering 
\subfloat[]{%
  \includegraphics[width=0.45\textwidth]{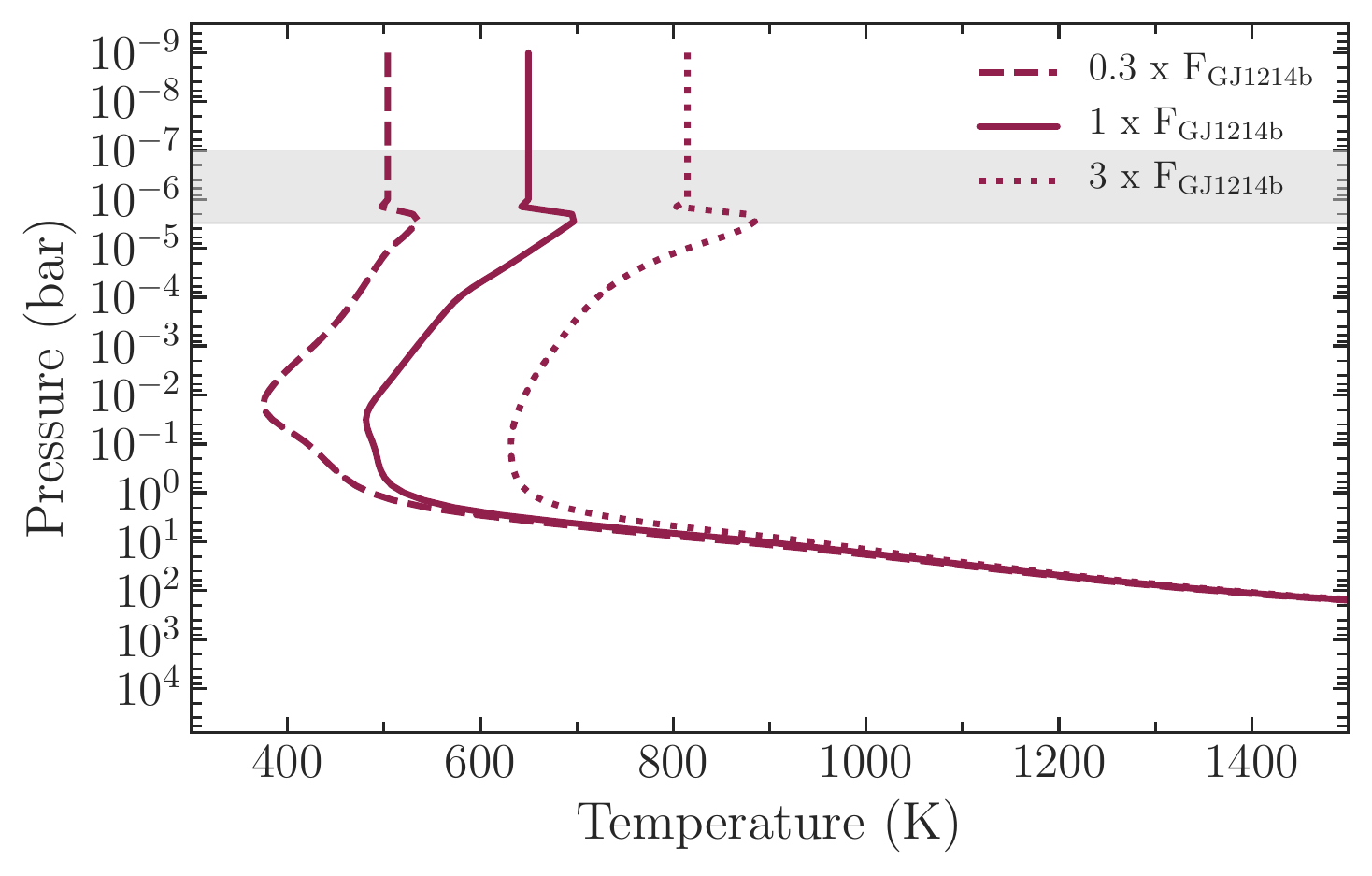}%
  \label{plot:PTprofs}%
}
\subfloat[]{%
  \includegraphics[width=0.45\textwidth]{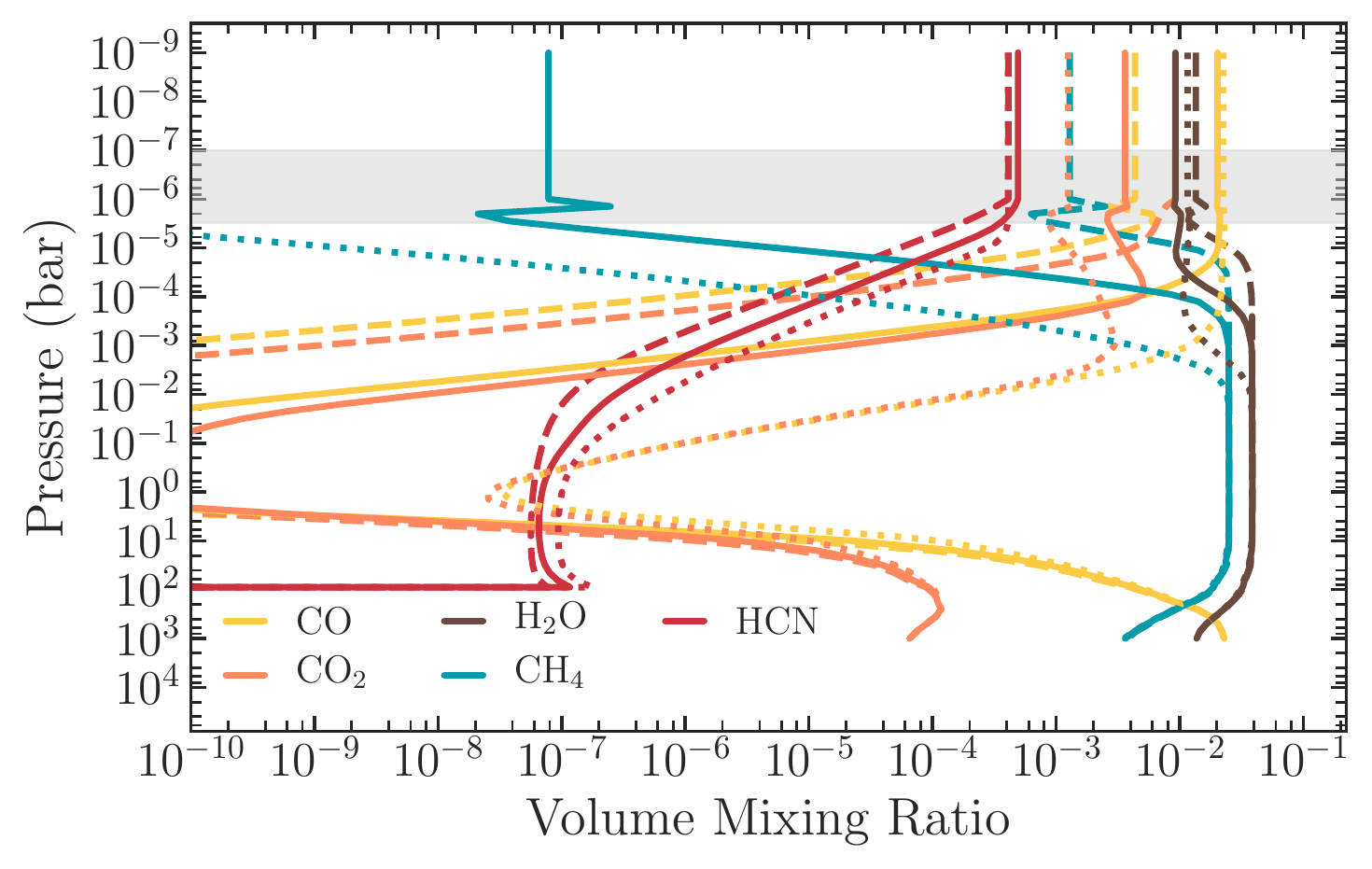}%
  \label{plot:AbundsProfs}%
}
\caption{Pressure-temperature profiles and molecular abundances for models with 50$\times$ solar metallicity and 0.3$\times$ (dashed), 1$\times$ (solid), and 3$\times$ (dotted) GJ 1214b's insolation. The high-resolution observations discussed in this paper are most sensitive roughly between 3 $\times$ 10$^{-6}$ - 3 $\times$ 10$^{-7}$ bars, marked by the light grey region in each plot.} 
\label{fig:Profiles}
\end{figure*}

\indent We generate high-resolution transmission spectra based on the 1D radiative-convective-photochemical models presented in \cite{Morley2015}. The authors assume 50$\times$ solar metallicity and use a 1D radiative-convective model to determine a temperature-pressure profile for the atmosphere assuming radiative-convective equilibrium and calculate gas abundances in different layers of the atmosphere assuming chemical equilibrium. They find the total mass of soot precursors from a photochemical model (\citealp{Kempton2012}; results first published in \citealp{Fortney2013}) at each layer and assume some percentage (f$_{\text{haze}}$) will form a scattering haze at that layer, with f$_{\text{haze}}$ and the mode particle size as free parameters. The optical properties of this haze are calculated with Mie theory. For this study, we focus on a particular combination of parameters that reproduce the ``flat'' \cite{Kreidberg2014} observations at low-resolution: f$_{\text{haze}}$ = 10\% and a mode particle size of 0.1 microns. This haze becomes opaque at a pressure of approximately 10$^{-5}$ bar in the atmosphere. We also look at models for atmospheres with 0.3$\times$ and 3$\times$ the insolation of GJ 1214 b with the same haze parameters but not the exact same haze. The atmospheres with 0.3$\times$, 1$\times$, and 3$\times$ the insolation of GJ 1214 b have effective temperatures of 412, 557, and 733 K, respectively \citep{Fortney2013}. The atmosphere models from \citet{Morley2015} go to 10$^{-6}$ bars at the top of the atmosphere, but our highest resolution spectra are sensitive out to $\sim 3 \times 10^{-7}$ bars, so we assume an isothermal atmosphere above $10^{-6}$ bars with constant molecular abundances.  The resulting pressure-temperature profiles and molecular abundances we use are shown in Figure \ref{fig:Profiles}.

\begin{figure}[htbp]
\centering
\includegraphics[width=.98\linewidth]{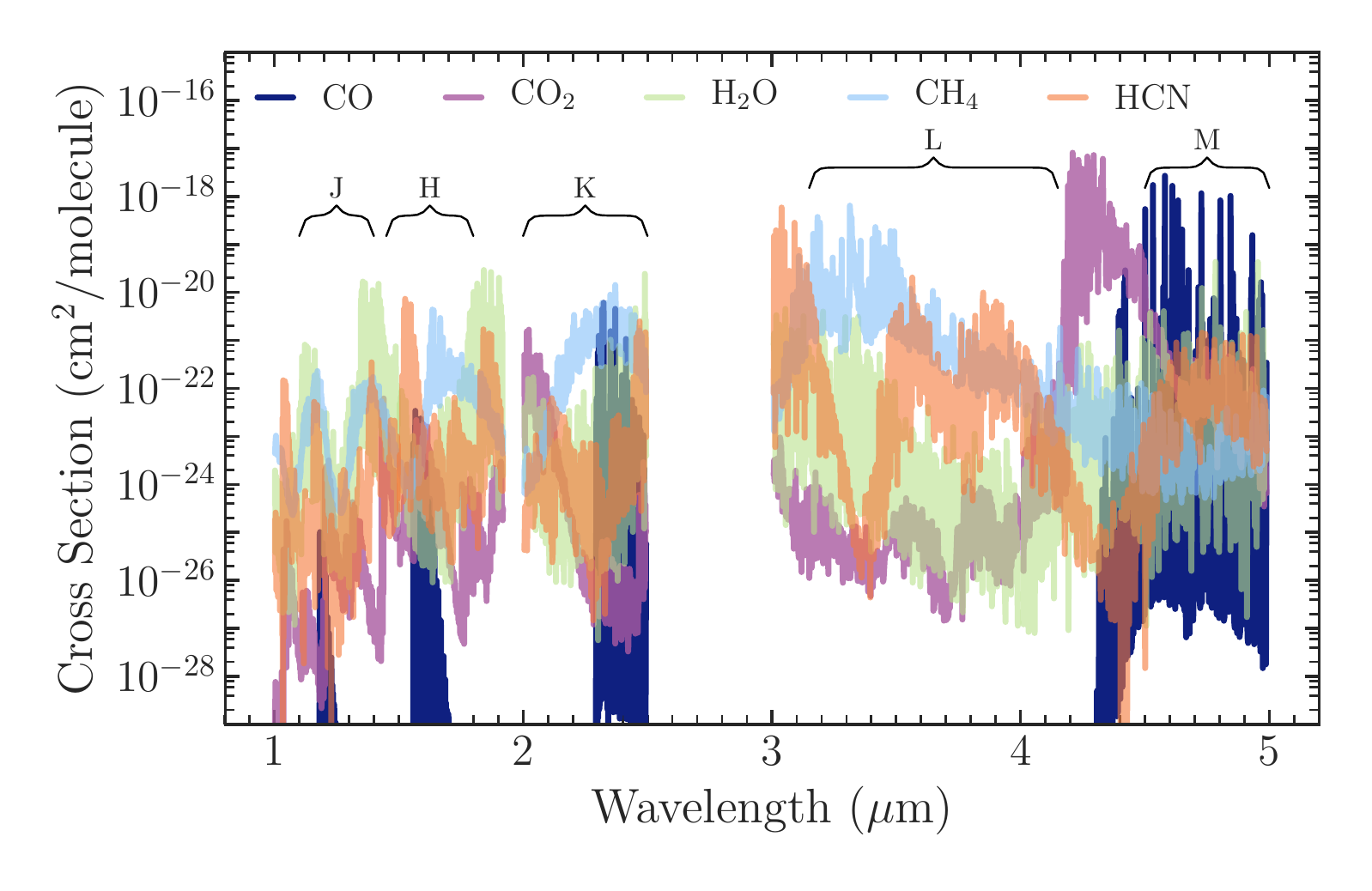}
  		\caption{Absorption cross sections for the molecules included in our spectra. These cross sections are calculated at a pressure of 10$^{-6}$ bars and a temperature of 650 K, then smoothed to R $\sim$ 1000 for illustrative purposes. Molecules with the strongest features in a particular bandpass, e.g. CO in M band, will be the dominant species in that wavelength range, though this effect is also dependent on the abundance of the molecule (see Figure \ref{plot:AbundsProfs}).}
  		\label{fig:xsecs}
\end{figure}

\indent To produce high-resolution transmission spectra, we use the flexible radiative transfer code described in the appendix of \cite{Morley2017}. This line-by-line code takes in the temperature-pressure profiles, chemical abundance profiles, and haze opacity files from \citep{Morley2015}, as well as a mass and radius of the planet. Using the the line-by-line optical depth calculations and the vectorized method for calculating transmission spectra presented in \citep{Robinson2017}, this code outputs high resolution line-by-line ($R\sim$500,000) transmission spectra for the planet. We use the cross-section database described in \citep{Freedman2014}.  For our calculations we include the opacities of CO \citep{Rothman2010}, CO$_{2}$ \citep{Huang2014,Huang2013}, H$_{2}$O \citep{Barber2006}, CH$_{4}$ \citep{Yurchenko2014,Yurchenko2013}, HCN \citep{Harris2008}, and H$_2$/He collision-induced absorption \citep{Richard2012}. Absorption cross sections for these molecules are shown in Figure \ref{fig:xsecs}. A molecule will be easiest to detect where it has the highest cross sections/strongest spectral features, for example H$_{2}$O in \textit{J} and \textit{H} bands, CH$_{4}$ in \textit{L} band, and CO in \textit{M} band.  \footnote{Although there have been updates to line lists for certain species since these publications, we do not include them in this study as we are not comparing to observations and are thus internally consistent.} 

\begin{figure}[htbp]
\centering
\includegraphics[width=.97\linewidth]{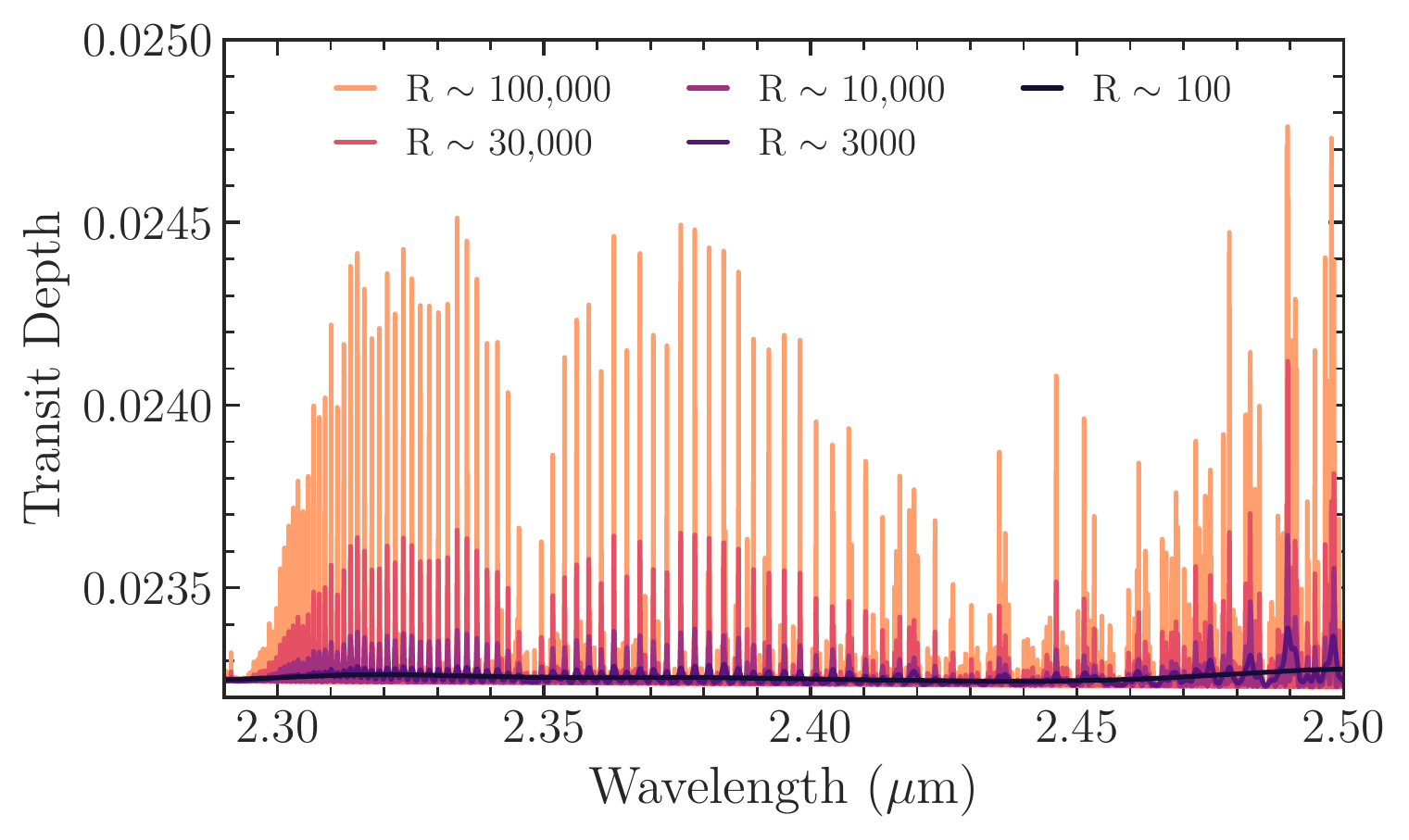}
  		\caption{Transmission spectra for a GJ 1214b model at the nominal insolation level across a range of spectral resolutions. At low resolution, the transmission spectrum is essentially a flat line, but spectral features are clearly visible at higher resolutions.}
  		\label{fig:resdemo}

\end{figure}

\begin{figure*}[tp]
\caption{Clear and hazy transmission spectra for models with 0.3$\times$ (top), 1$\times$ (middle) and 3$\times$ (bottom) GJ 1214b's insolation. Both low (R $\sim 100$) and high (R $\sim 100,000$) resolution spectra are plotted. The presence of the haze clearly mutes molecular features, particularly in the low resolution spectra. Furthermore, at high resolution the difference between the coolest and hottest models is most pronounced; this difference can be attributed to the larger scale height for the hotter model in addition to differences in abundances.}
\centering

\includegraphics[width=.8\textwidth]{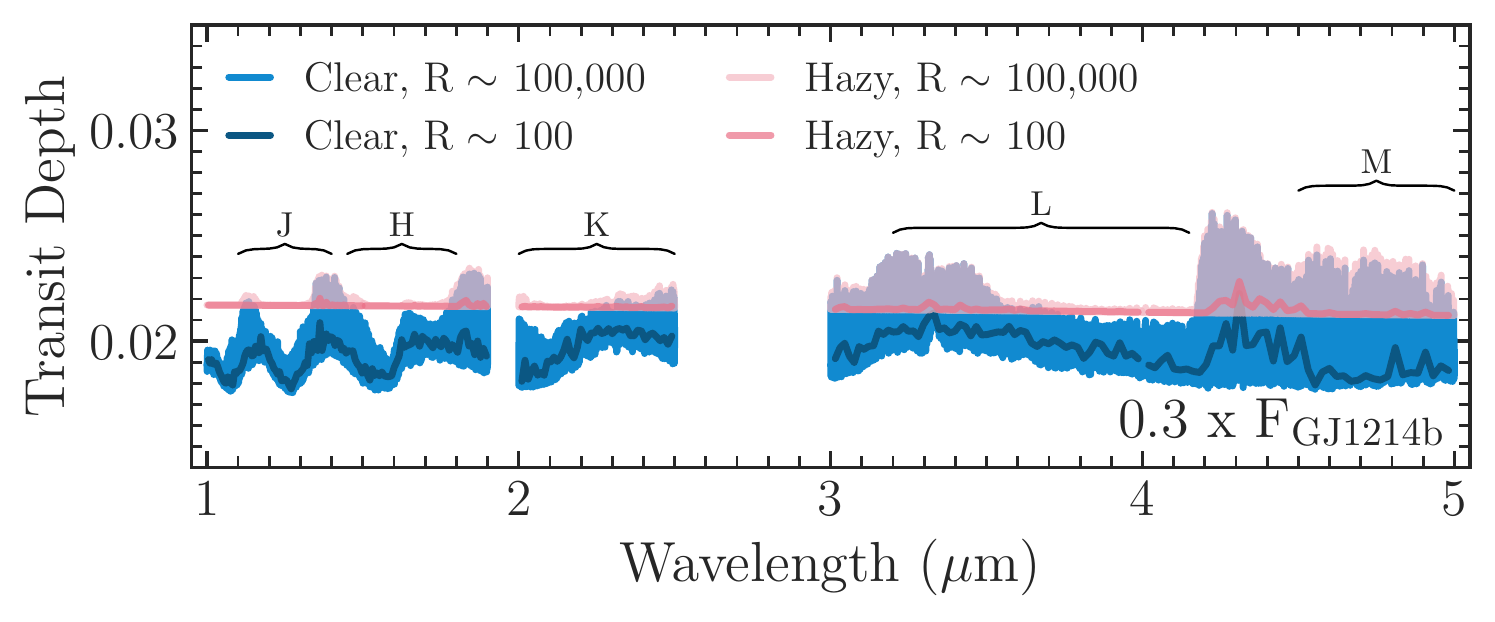}

\includegraphics[width=.8\textwidth]{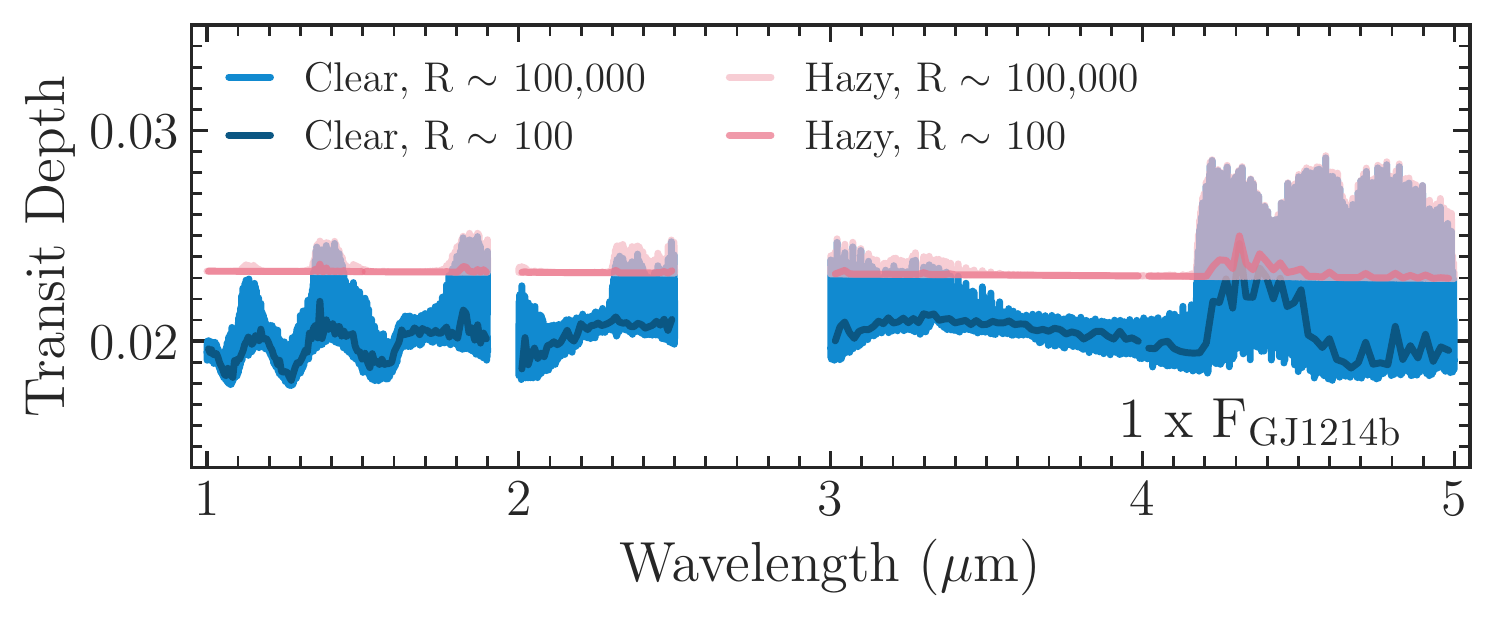}
\includegraphics[width=.8\textwidth]{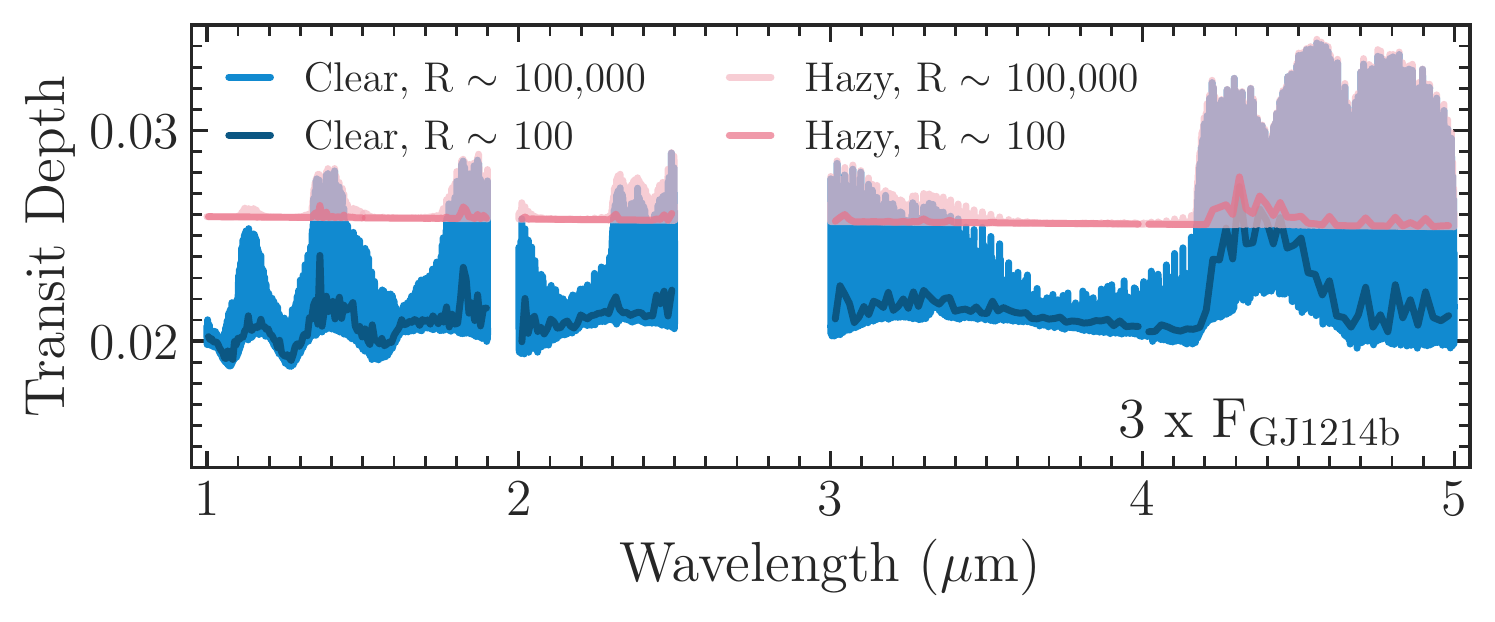}

\label{fullspecs}
\end{figure*}

Example transmission spectra in \textit{K} band for a range of resolutions are shown in Figure \ref{fig:resdemo}. Though the low resolution transmission spectrum shows little deviation from a flat line, resolved spectral features from CO and H$_{2}$O are visible starting with R$\sim$10,000 and increase in size as spectral resolution increases. High and low resolution spectra from 1 to 5 $\mu$m with and without the haze for all three insolation cases are shown in Figure \ref{fullspecs}. The haze opacity effectively obscures the molecular features below a certain pressure in the atmosphere, reducing the low resolution spectrum in particular to a mostly flat line. As the stellar insolation (and therefore the effective temperature of the atmosphere) increases, CH$_{4}$ features in L band disappear while CO shows stronger features in K and M band, in accordance with the change in abundances shown in Figure \ref{plot:AbundsProfs}.

\begin{figure}[htbp]
\centering
\includegraphics[width=.95\linewidth]{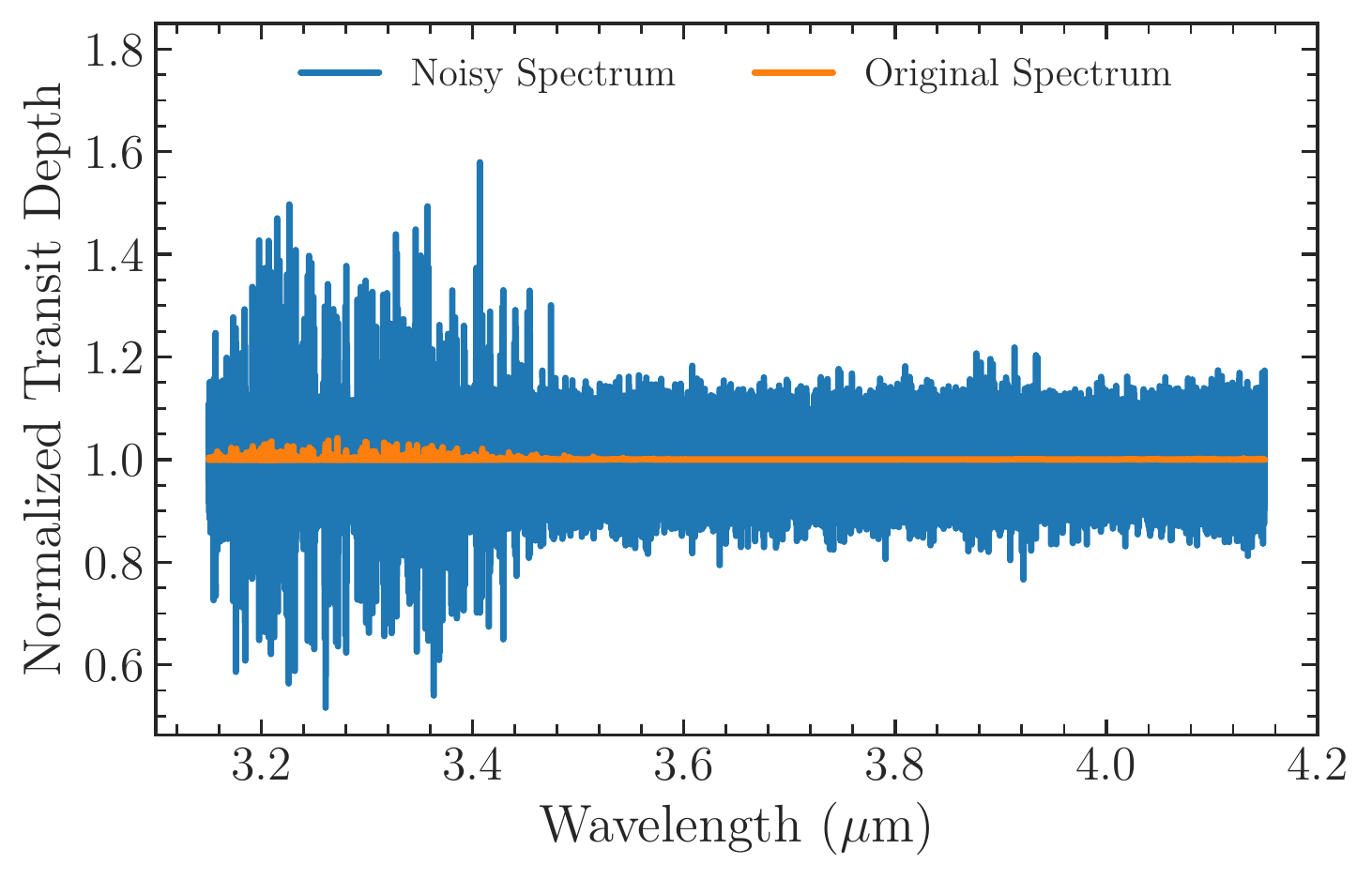}
  		\caption{Normalized transmission spectra with R $\sim$ 100,000 for a GJ 1214b model before and after the addition of noise. The random noise added to the spectrum has a base S/N$_{res}$ of 1000 (or 1000 ppm of noise), scaled by a model of the Earth's transmission spectrum.}
  		\label{fig:noisedemo}

\end{figure}

\indent Previous analyses of high resolution spectra have involved various methods to remove telluric and stellar contamination of the data which all require normalization of the observed spectra. To approximate an observed and reduced spectrum, we take a $R\sim$500,000 transmission spectrum model including all opacity sources as described above, Doppler shift it according to the systemic velocity of GJ 1214 (21 km/s), and then to reach the desired spectral resolution smooth with a Gaussian kernel and interpolate the model onto a coarser wavelength grid corresponding to the given resolution and 2 pixels per resolution element (assuming a Nyquist like sampling of 2 pixels per element).  We assume  telluric  and  stellar contamination will be dealt with completely by the data reduction process such that they are removed to the photon noise level. Consequently, we only simulate the planet’s transmission spectrum, and investigate the effect of the photon noise. While we do not assume a full noise simulator and data processing steps \citep[e.g.][]{Brogi2019}, we approximate the final outcome of such an approach by doing the following. First, to determine the amount of random noise to add to each pixel in our “truth” spectrum, we choose a particular signal-to-noise ratio on what would have been observed in the stellar spectrum per resolution element (S/N$_{\text{res}}$) as in \citet{Pino2018b}.  For the signal-to-noise per pixel, S/N$_{\text{pix}}(\lambda)$ we multiply this S/N$_{\text{res}}$ by the square root of the telluric absorption spectrum T($\lambda$) (which has values between 0 and 1), to mimic the reduction in S/N due to telluric extinction, and the square root of the number of pixels per resolution element. We simulate noise by adding a noise value to each pixel drawn from a Gaussian distribution with a standard deviation equal to the reciprocal of the desired signal-to-noise per pixel. Given that we are assuming photon noise, which follows a Poisson distribution, Gaussian distributed noise is an appropriate approximation for the high stellar photon counts. We obtain T($\lambda$) using the ESO Skycalc tool based on the Cerro Paranal Sky Model \citep{Noll2012, Jones2013}. An example of the normalized transmission spectrum in \textit{L} band before and after the addition of noise is shown in Figure \ref{fig:noisedemo}. Both the original GJ 1214b transmission spectrum and that of Earth's atmosphere have stronger features on the bluer end of L band, reflected in the noisier but more prominent features on the shorter wavelength end of the noisy spectrum.

\subsection{Quantifying Detection Significance}

\indent High-resolution spectra of exoplanet atmospheres are often ``self-calibrated" rather than in comparison to a standard star, meaning broadband information and changes in flux at a fixed wavelength over time are removed from the data by fitting a trend with airmass or using a principal component analysis based approach \citep[e.g.][]{Snellen2010, Birkby2013}.  These data processing steps to remove the telluric contamination typically remove any reliable planetary continuum level information. This makes typical data-model ``chi-square" comparisons difficult, if not impossible, unless the exact stretching/scaling to the data is known.  In light of this, the standard approach is to utilize the cross-correlation function (CCF), which leverages information in individual line ratios to determine planetary atmosphere information (see \citealt{Brogi2019} and \citealt{Gibson2020} for a detailed discussion).  The cross-correlation function determines the correlation between the data and a model template as a function of Doppler shift.  A model perfectly matched to the data will show a ``peak", or maximum correlation, at the planetary velocity (in our case, the systemic velocity).  Incorrect models will show no peak or a damped peak relative to the ``correct" model. Following the notation of \citet{Brogi2019}, we define the variance of the data (s$^{2}_{ f}$), the variance of the model (s$^{2}_{ g}$ ), and the cross-covariance R(s) as follows:

\begin{equation*}
    s^{2}_{ f} = \frac{1}{N} \sum\limits_{n} f^{2}(n) \\
\end{equation*}
\begin{equation*}
    s^{2}_{ g} = \frac{1}{N} \sum\limits_{n} g^{2}(n-s) \\
\end{equation*}
\begin{equation*}
     R(s) = \frac{1}{N}  \sum\limits_{n} f(n) g(n-s)
\end{equation*}
where n is the bin number or spectral channel, s is a bin/wavelength shift, N is the total number of spectral channels, f(n) is an observed spectrum, and g(n) is a model spectrum for comparison, both mean-subtracted.

The cross-correlation coefficient C(s) is then:
\begin{equation}
    C(s) = \frac{R(s)}{\sqrt{s^{2}_{f} s ^{2}_{g}}}
\end{equation}

\indent In the literature, molecules have been detected by reporting a strong signal in the cross-correlation function of observed spectra with a model that contains solely the molecule of interest. When using the CCF, we compare to models that have only one or two opacity sources at a time, e.g. just CO or CO and the haze opacity. The strength of this detection has often been reported as the ratio of the peak of the CCF and the standard deviation of the coefficient around the peak \citep[e.g.][]{Snellen2010} though some more sophisticated approaches have also been used like the Welch T-test metric \citep[e.g.][]{Birkby2013}. \citet{Hawker2018} find the ratio of peak to standard deviation to be the most conservative metric for evaluating detection significance, so this is how we will report detection significances from a CCF in this work.  However, this peak-to-off-peak comparison only determines the S/N within a given model template relative to the on-to-off velocities, making quantitative comparisons amongst differing model templates challenging. 

\indent \citet{Brogi2019} proposed a solution to this problem by developing a mapping of the CCF to a log-likelihood function (log(L)) for use in a Bayesian retrieval framework. Using the above definitions, they related a formal log-likelihood to the CCF (or rather, the cross-covariance) through:
\begin{equation}\label{eq:logl}
    \text{log(L)} = -\frac{N}{2}\log(s^{2}_{f} - 2 R(s) + s ^{2}_{g})
\end{equation}
where as defined above N is the total number of spectral channels, s$^{2}_{ f}$ is the variance of the data, s$^{2}_{ g}$ is the variance of the model, and R(s) is the cross-covariance of the data and a model with some wavelength shift s. Before proceeding with our atmospheric analysis, we first investigate the sensitivity of the model comparisons under the CCF and log(L) assumptions. Since we are not performing a retrieval analysis but instead comparing to forward models, we need a method of mapping the log(L) value to a detection significance analogous to that obtained with the CCF. To do this, we first calculate log($L_{1}$) for our ``truth'' spectrum (e.g., the same underlying model used to generate the simulated data, which again, includes the haze continuum and all of the gases). We then compute log($L_{2}$) for 5 additional nested models that each lack one of our tested opacity sources, so we can isolate how much that missing opacity source decreases the log(L). To quantify the detection of each source of opacity, we utilize the change in Bayesian Information Criterion (BIC) \citep{Schwarz1978}:
\begin{equation}\label{eq:BIC}
    \text{BIC} = p \log{N} - 2\log{L}
\end{equation}
between the full model and the subset model lacking that opacity source, of which we then approximate $\Delta_{BIC} \approx 2* (\log{L_{1}}-\log{L_{2}})$.\footnote{Effectively, this becomes a likelihood ratio test to compare models as we are not changing the number of free parameters (p in Equation \ref{eq:BIC}). The molecule we remove is not a free parameter as we do not vary its value to fit the data.}. We then relate this change in BIC to the Bayes factor, using the formula $\Delta_{BIC} = 2*\log{B_{12}} $ \citep[e.g.][]{Szyd2015}. We can then map this Bayes factor to a frequentist p-value using Table 2 from \citet{Trotta2008}, which can in turn be converted into a statistical significance. Due to the limited nature of the table, we are only able to report signficances smaller than 21.3$\sigma$; anything that would have a stronger significance is reported as this upper limit.

\subsection{CCF vs. Log(L) Example: CO in \textit{K} Band}\label{subsec:COex}
Here we compare molecular detections from the CCF and log(L) approaches as a function of the signal-to-noise per resolution element S/N$_{res}$ for a representative \textit{K}-band (2-2.5 $\mu$m) spectrum of our nominal GJ 1214b model.  

\begin{figure}[htbp]
\centering
\includegraphics[width=.95\linewidth]{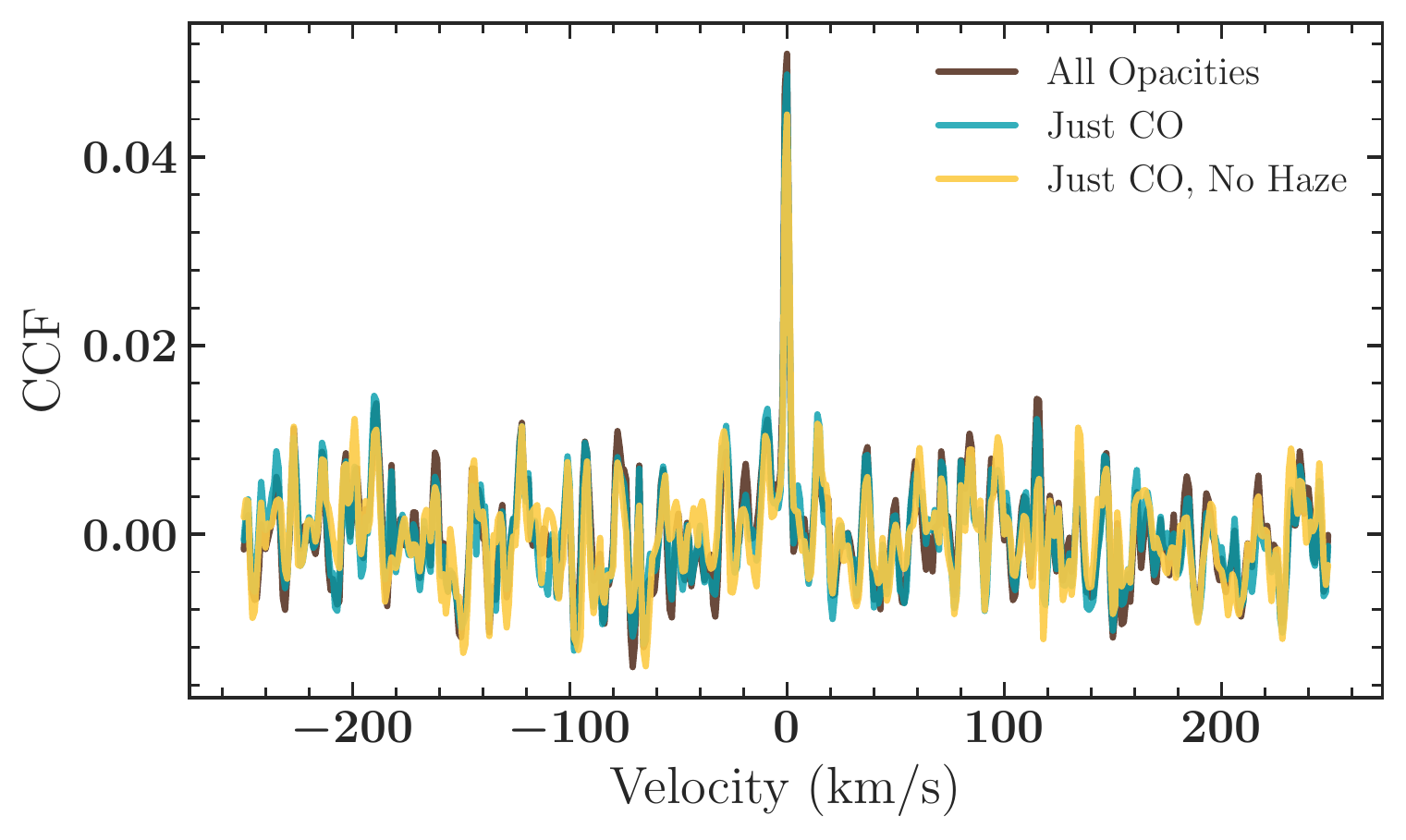}
  		\caption{CCFs of transmission spectrum models that include varying opacity sources with a spectrum that includes all opacities and has random noise added to give it a S/N per resolution element of 1000.}
  		\label{fig:CCFdemo}

\end{figure}

\indent First, we look at a single truth spectrum with S/N$_{res}$ of 1000 and R$\sim$100,000, assuming a velocity of 0 km/s for the planet. Figure \ref{fig:CCFdemo} shows the CCFs of this spectrum with models that contain either all opacity sources, just CO and the haze, or just CO; the statistical significances of each peak are 9.9$\sigma$, 9.7$\sigma$, and 9.1$\sigma$, respectively \footnote{Typically in the literature this metric is referred to as the S/N of a detection and lacks the sigma symbol (unlike significance values from the Welch T-test metric for example). However, we will include the sigma in this work to aid in comparison to detection significances derived from the log(L) method.
}. As expected, the template that contains all of the opacity sources included in the original model gives the highest peak CCF, relative to the off-peak velocity baseline. The template that contains both CO and the haze has a slightly smaller peak, which again decreases when the haze opacity is ignored.  However, these decreases are relatively small, suggesting the CCF is not particularly sensitive to the presence of a haze. 

\begin{figure}
\centering 
\subfloat[]{%
  \includegraphics[width=0.95\columnwidth]{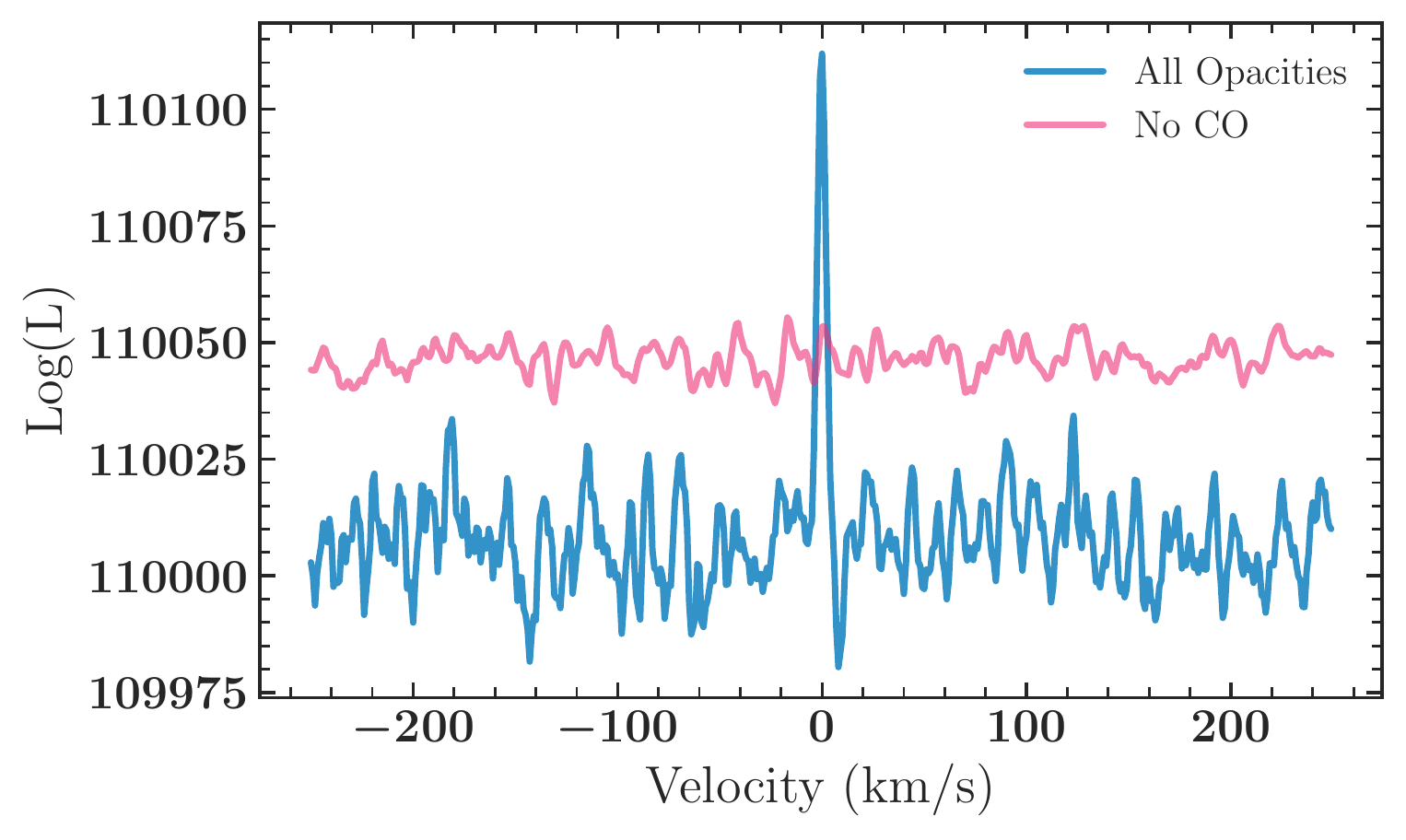}%
  \label{plot:LogLzoom}%
}\qquad
\subfloat[]{%
  \includegraphics[width=0.95\columnwidth]{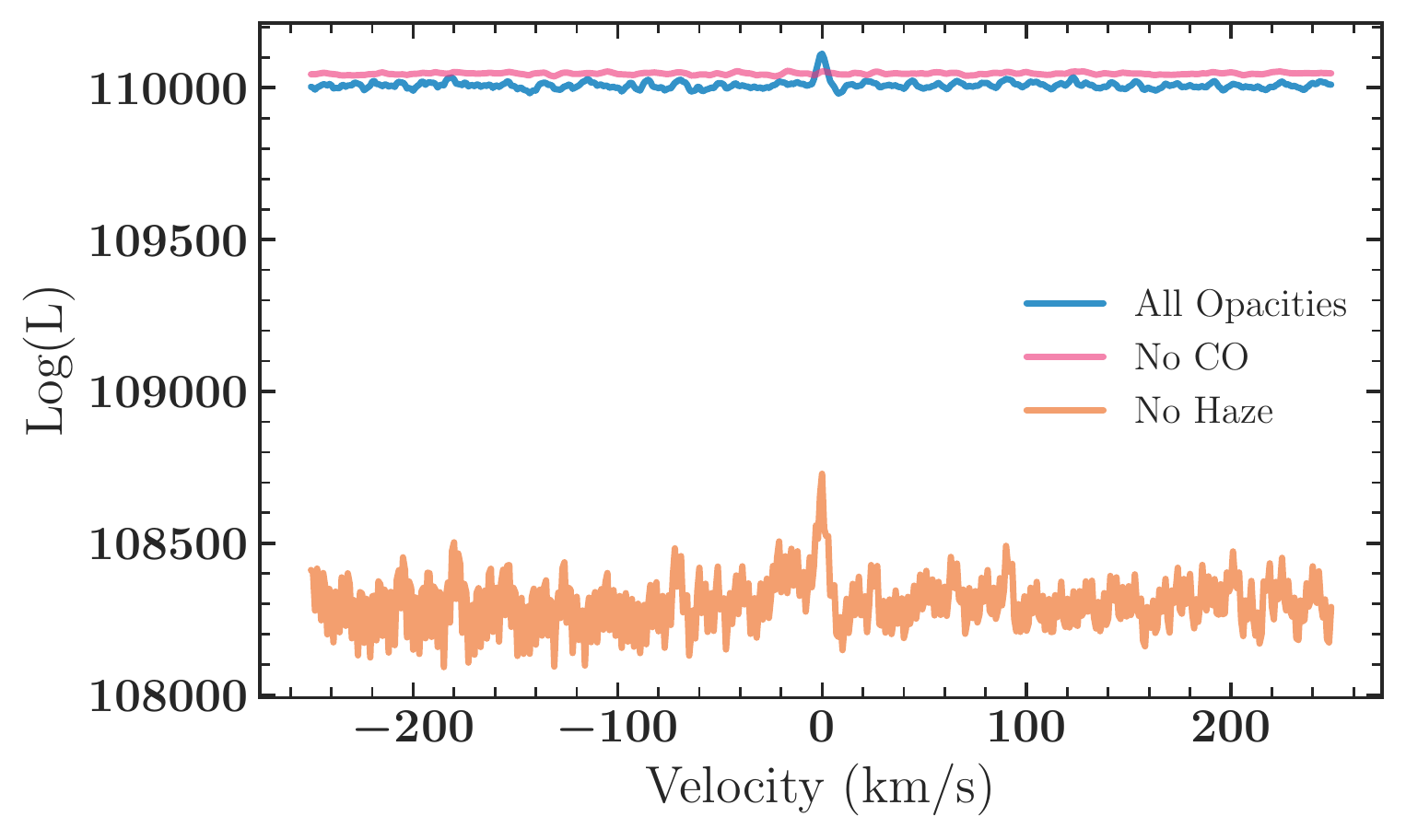}%
  \label{plot:LogLfull}%
}
\caption{Log(L) for transmission spectrum models that include varying opacity sources with a spectrum that includes all opacities and has random noise added to give it a S/N per resolution element of 1000. In \ref{plot:LogLzoom}, we see that the peak in log(L) seen at zero relative velocity disappears when CO is removed from the model. In \ref{plot:LogLfull}, we see the value of log(L) drastically decreases when the haze opacity is removed from the model.}
\label{fig:LogLdemo}
\end{figure}

\indent Figure \ref{fig:LogLdemo} similarly shows the log(L) as a function of velocity for templates that include all opacities, all but CO, and all but the haze. Figure \ref{plot:LogLzoom} is zoomed in to show how the peak seen when computing log(L) for the full model disappears when CO is removed, highlighting the necessity of CO to properly match the original spectrum. The decrease in log(L) at the planet velocity (0 km/s) corresponds to a  10.9 $\sigma$ detection of CO (based on Eqn. \ref{eq:BIC} and subsequent discussion). We note that the log(L) for the model without CO has a higher average value than the model with CO; we attribute this effect to the decreased variance ($s_{g}$) of the model without the prominent CO lines, which increases the resulting log(L) as seen in Equation \ref{eq:logl}. Thus, a clear peak in the log(L) as a function of velocity should be taken as an indication that a model that correctly matches the truth spectrum rather than just the average value. We assume a retrieval method would identify the correct velocities and so only consider the value at 0 km/s as this is most analogous to what a retrieval detection significance would be. Figure \ref{plot:LogLfull} shows the same curves, as well as log(L) with a template not containing the haze. When the haze opacity is removed, though the peak in log(L) at 0 km/s is still prominent, log(L) significantly decreases at this velocity, corresponding to a $>20\sigma$ detection of the haze. Therefore, the haze is also necessary in addition to CO to match the truth spectrum. However, if one did not test the models containing the haze opacity, they may still detect CO since the Log(L) still shows a peak at 0 km/s but assume the atmosphere was clear. Models including a potential source of continuum opacity should be investigated to maximize the atmospheric information one can learn from the data.

\begin{figure*}
\centering 
\subfloat[CCF: CO and Haze]{%
  \includegraphics[width=0.45\textwidth]{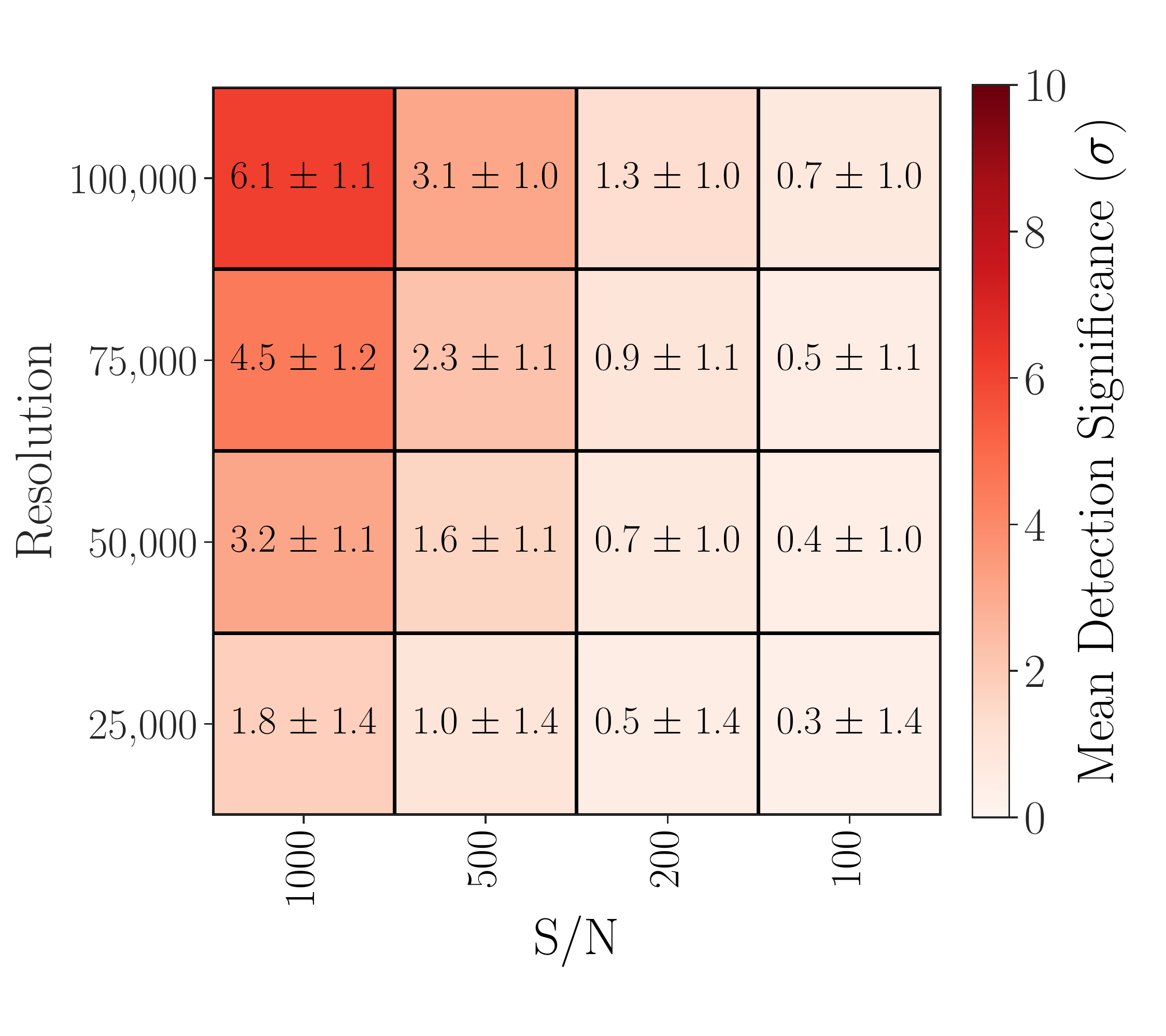}%
  \label{plot:CCFcolormap_CO_haze}%
}\qquad
\subfloat[CCF: Just CO]{%
  \includegraphics[width=0.45\textwidth]{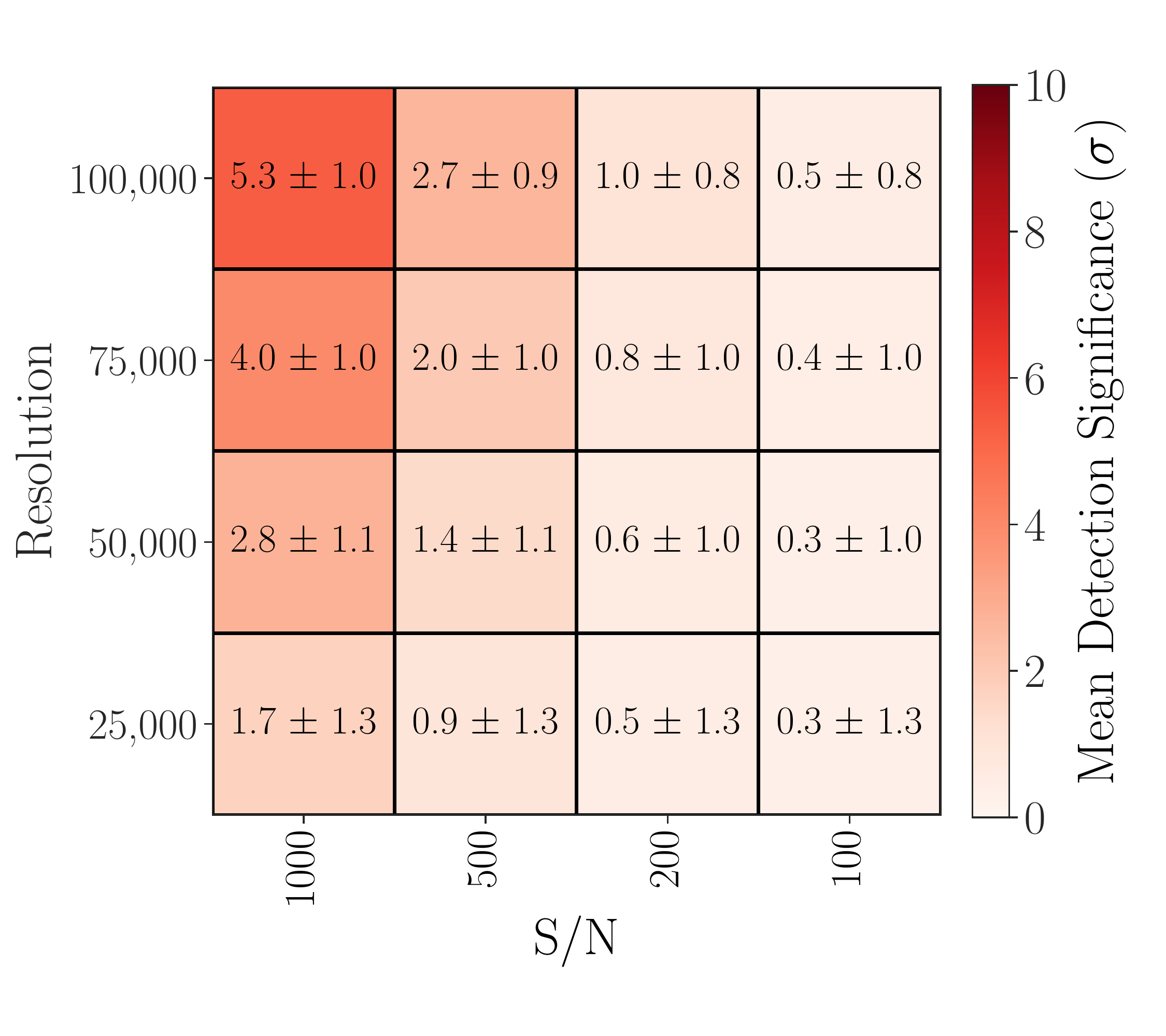}%
  \label{plot:CCFcolormap_CO_nohaze}%
}
\caption{Detection significance in sigma for CO as a function of spectral resolution and S/N per resolution element of the transmission spectrum of nominal GJ 1214b model (1$\times$ insolation). In \ref{plot:CCFcolormap_CO_haze}, we have cross-correlated our input spectra with models including opacity only from CO and a haze. In \ref{plot:CCFcolormap_CO_nohaze}, we have removed the haze opacity from the model we use for cross-correlation; the detection significances agree with those obtained when including the haze opacity, making it difficult to robustly identify the presence of a haze when using the CCF.}
\label{fig:CCFcolormaps}
\end{figure*}

\begin{figure*}
\centering 
\subfloat[Log(L): CO]{%
  \includegraphics[width=0.45\textwidth]{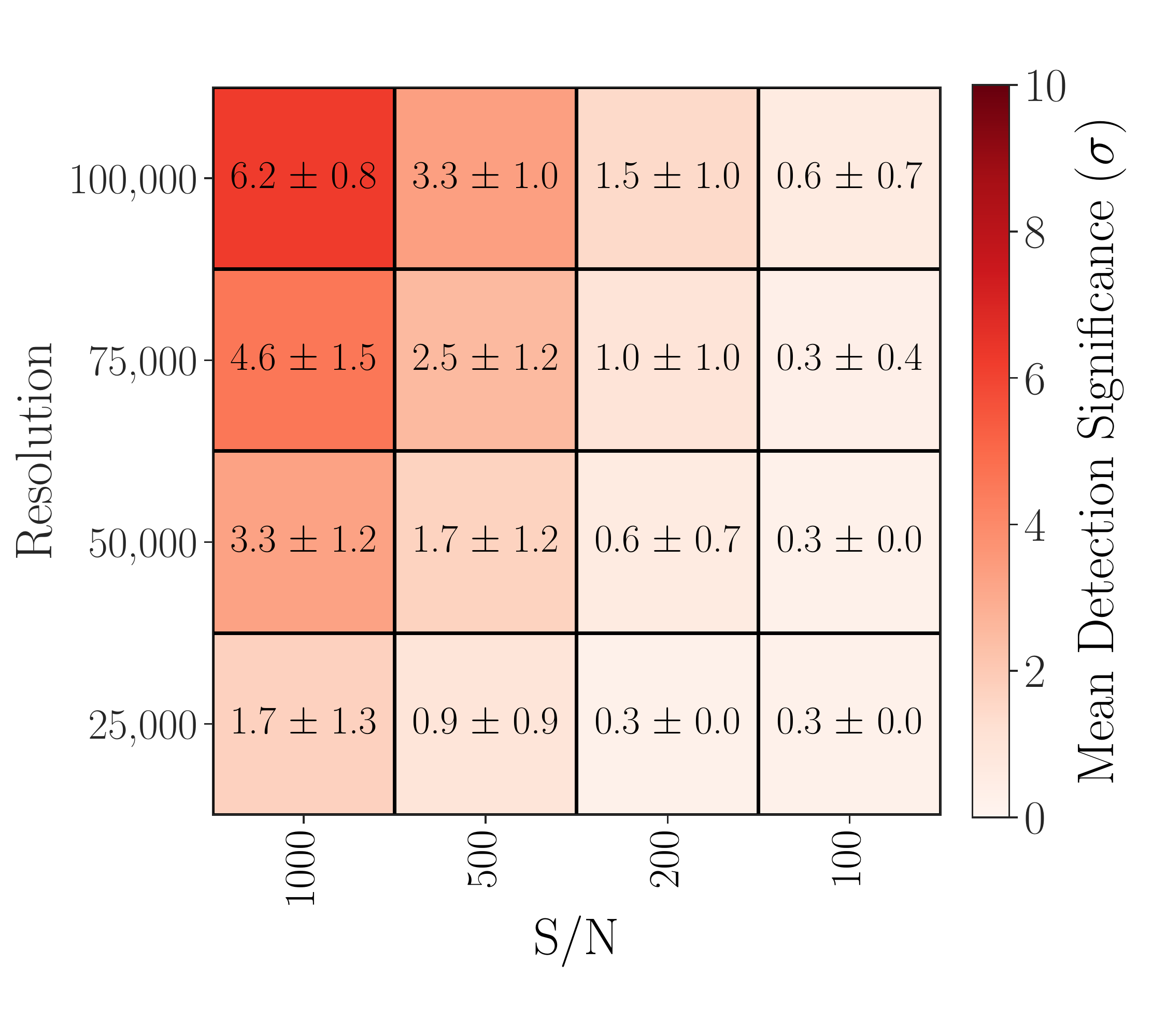}%
  \label{plot:LogLcolormap_CO}%
}\qquad
\subfloat[Log(L): Haze]{%
  \includegraphics[width=0.45\textwidth]{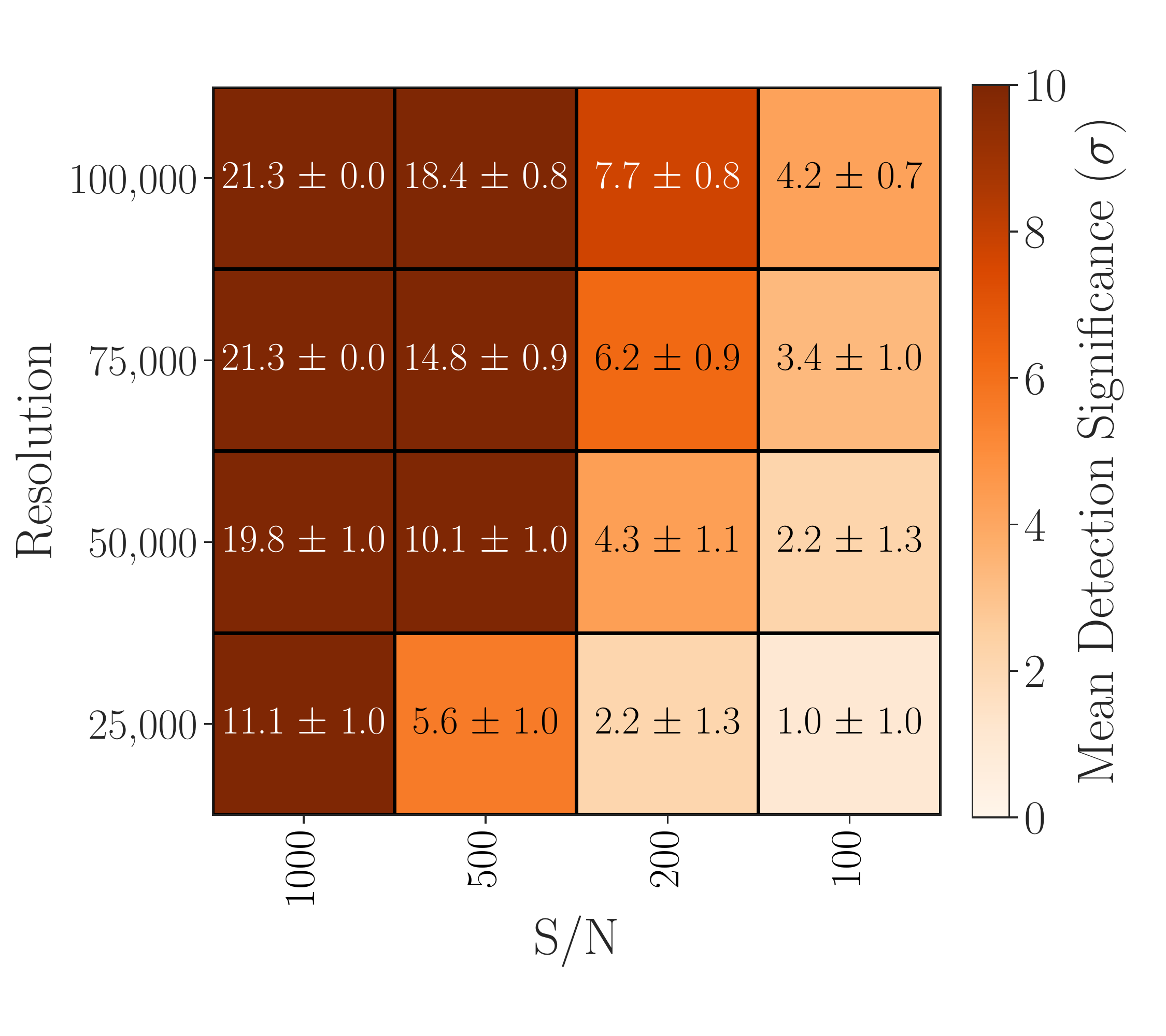}%
  \label{plot:LogLcolormap_haze}%
}
\caption{Detection significances in sigma for CO and a haze as a function of spectral resolution and S/N per resolution element of the transmission spectrum of nominal GJ 1214b model (1$\times$ insolation). The detection strengths reported in \ref{plot:LogLcolormap_CO} are very similar to those found when detecting CO with the CCF as shown in \ref{plot:CCFcolormap_CO_haze}. However, as this method allows us to probe the presence of a haze directly, we can much more confidently report the detection of a haze using the log(L) as shown in \ref{plot:LogLcolormap_haze}.}
\label{fig:LogLcolormaps}
\end{figure*}

\indent Next we explore the differences in detection from the CCF vs. log(L) approach over a small grid of S/N$_{res}$ and resolutions, summarized in Figures \ref{fig:CCFcolormaps} and \ref{fig:LogLcolormaps}. In order to account for the effect of random noise, we repeat the analysis for 25 different noise instances and report the average and standard deviation of the resulting detection strengths. Comparing Figures \ref{plot:CCFcolormap_CO_haze} and \ref{plot:CCFcolormap_CO_nohaze}, we see that removing the haze opacity does typically slightly lower the average detection strength of CO when using the CCF regardless of spectral resolution or S/N$_{res}$. However, within their uncertainties, the detection strengths of CO agree between templates that include or ignore the haze opacity.  Figures \ref{plot:LogLcolormap_CO} and \ref{plot:LogLcolormap_haze} show the detection strengths for CO and haze, respectively, when using the log(L) method. We see that \ref{plot:LogLcolormap_CO} resembles \ref{plot:CCFcolormap_CO_haze} and \ref{plot:CCFcolormap_CO_nohaze}, meaning that the CCF and log(L) give similar detection strengths for CO. Figure \ref{plot:LogLcolormap_haze} shows we can strongly detect the presence of a haze, even when we cannot robustly detect CO (for example, with R$\sim$ 25,000). 

\indent Overall, we find that CCF and log(L) give similar answers for molecular detections, but the log(L) method is much more sensitive to the presence of a haze opacity. This suggests the high-resolution spectrum is sensitive to the broadband opacity of the haze due to the loss of a myriad of weaker lines. The increased sensitivity of the log(L) is attributable to the treatment of the model and data variance terms (s$_{f}$ and s$_{g}$). As noted in \citet{Brogi2019}, the log(L) decreases when s$_{f}$ and s$_{g}$ differ significantly, while the CCF is not clearly affected by this discrepancy. The presence of a haze opacity will serve to mute any molecular features (as seen in Figure \ref{fullspecs}), effectively decreasing the variance of the spectrum. As a result, a model that includes the haze will have a more similar variance to the observed spectrum of a hazy object, leading to a detectable change in log(L).  Thus, for the remainder of this paper, we will only use the log(L) method to quantify how well we can detect opacity sources.

\section{Results} \label{sec:results}
Here we present the detectability given the above metrics for CO, CO$_{2}$, H$_{2}$O, CH$_{4}$, and a haze opacity as a function of spectral band, resolution, and S/N$_{res}$. The ``truth" spectrum is generated by including all of the above opacity sources as well as H$_{2}$/He collision-induced absorption from one of three input pressure-temperature profiles, corresponding to 0.3$\times$, 1$\times$, or 3$\times$ GJ 1214 b's stellar insolation (Figure \ref{fig:Profiles}). The observing bands we consider and their corresponding wavelengths are shown in Table \ref{tab:obsbands}. These bandpasses are a bit wider than are typically defined for ground based observations as at high resolution telluric absorption features can possibly be resolved with usable data in between, and instruments can have varying wavelength coverage. We test a large range of S/N$_{res}$ from 50 to 5000, averaging the detectability over 25 random noise instances (as in Section \ref{subsec:COex}). We then find the average detection strength from these 25 noise instances. Tables \ref{tab:CO} - \ref{tab:Haze} report the lowest S/N$_{res}$ required for detecting (threshold for detection set at 5$\sigma$) a given opacity source (one table per source) as a function of observing band, insolation (relative to that of GJ 1214b), and spectral resolution. We will go into a more detailed overview of these results in the following sections.

\begin{deluxetable}{c|c}\label{tab:obsbands}
    \tablecolumns{2}
    \tablewidth{1pt}
    \tablecaption{Observing bands considered in this study.}
    \tablehead{\colhead{Observing Band} & \colhead{Wavelength Coverage (microns)}}
    \startdata
    \textit{J} & 1.1 - 1.4 \\
    \textit{H} & 1.45 - 1.8 \\
    \textit{K} & 2.0 - 2.5 \\
    \textit{L} & 3.2 - 4.15 \\
    \textit{M} & 4.4 - 5.0 \\
    \enddata
\end{deluxetable}

\setlength{\tabcolsep}{5pt}
\begin{deluxetable}{cccccc}\label{tab:Haze}
    \tablecolumns{6}
    \tablewidth{1pt}
    \tablecaption{Minimum S/N$_{res}$ required for $\geq 5\sigma$ detection of the haze.}
    \tablehead{\colhead{Observing} & \colhead{Insolation} & \multicolumn{4}{c}{Spectral Resolution} \\
    \cline{3-6}
    \colhead{Band} & \nocolhead{} & \colhead{25,000} & \colhead{50,000} & \colhead{75,000} & \colhead {100,000} 
    }
    \colnumbers
    \startdata
    % J-band
     & 0.3$\times$ & 400 & 250 & 200 & 150 \\
    \textit{J} & 1$\times$ & 350 & 200 & 150 & 150  \\
     & 3$\times$ & 200 & 150 & 100 & 100  \\ [0.5 cm]
     % H-band
     & 0.3$\times$ & 350 & 200 & 150 & 150  \\
     \textit{H} & 1$\times$ & 350 & 200 & 150 & 100 \\
     & 3$\times$ & 200 & 100 & 100 & 100 \\ [0.5 cm]
     % K-band
     & 0.3$\times$ & 400 & 250 & 200 & 150  \\
     \textit{K} & 1$\times$ & 450 & 250 & 200 & 150 \\
     & 3$\times$ & 300 & 150 & 100 & 100 \\ [0.5 cm]
     % L-band
     & 0.3$\times$ & 300 & 200 & 150 & 100  \\
    \textit{L} & 1$\times$ & 400 & 250 & 200 & 150  \\
     & 3$\times$ & 300 & 200 & 150 &  100 \\ [0.5 cm]
     % M-band
     & 0.3$\times$ & 350 & 200 & 150 & 150  \\
     \textit{M} & 1$\times$ & 250 & 150 & 100 & 100 \\
     & 3$\times$ & 200 & 150 & 100 & 100 \\ [0.5 cm]
    \enddata
\end{deluxetable}

\setlength{\tabcolsep}{5pt}
\begin{deluxetable}{cccccc}\label{tab:CO}
    \tablecolumns{6}
    \tablewidth{0pt}
    \tablecaption{Minimum S/N$_{res}$ required for $\geq 5\sigma$ detection of CO. There are no $\geq 5\sigma$ detections with S/N$_{res}$ $\leq 5000$ for \textit{J}, \textit{H}, and \textit{L} bands.}
    \tablehead{\colhead{Observing} & \colhead{Insolation} & \multicolumn{4}{c}{Spectral Resolution} \\
    \cline{3-6}
    \colhead{Band} & \nocolhead{} & \colhead{25,000} & \colhead{50,000} & \colhead{75,000} & \colhead {100,000} 
    }
    \colnumbers
    \startdata
    % J-band
    % & 0.3$\times$ & - & - & - & - \\
    %J & 1$\times$ & - & - & - & -  \\
    % & 3$\times$ & - & - & - & -  \\ [0.5 cm]
     % H-band
    % & 0.3$\times$ & - & - & - & -  \\
    % H & 1$\times$ & - & - & - & - \\
    % & 3$\times$ & - & - & - & - \\ [0.5 cm]
     % K-band
     & 0.3$\times$ & - & 4500 & 3500 & 2400  \\
     \textit{K} & 1$\times$ & 3500 & 1600 & 1100 & 800 \\
     & 3$\times$ & 1800 & 1000 & 700 & 500 \\ [0.5 cm]
     % L-band
     %& 0.3$\times$ & - & - & - & -  \\
     %L & 1$\times$ & - & - & - & -  \\
     %& 3$\times$ & - & - & - &  - \\ [0.5 cm]
     % M-band
     & 0.3$\times$ & 1400 & 800 & 600 & 400 \\
     \textit{M} & 1$\times$ & 700 & 400 & 250 & 200  \\
     & 3$\times$ & 400 & 250 & 150 & 150  \\ [0.3 cm]
    \enddata
\end{deluxetable}

\setlength{\tabcolsep}{5pt}
\begin{deluxetable}{cccccc}\label{tab:CO2}
    \tablecolumns{6}
    \tablewidth{1pt}
    \tablecaption{Minimum S/N$_{res}$ required for $\geq 5\sigma$ detection of CO$_{2}$. There are no $\geq 5\sigma$ detections with S/N$_{res}$ $\leq 5000$ for \textit{J}, \textit{H}, \textit{K}, and \textit{L} bands.}
    \tablehead{\colhead{Observing} & \colhead{Insolation} & \multicolumn{4}{c}{Spectral Resolution} \\
    \cline{3-6}
    \colhead{Band} & \nocolhead{} & \colhead{25,000} & \colhead{50,000} & \colhead{75,000} & \colhead {100,000} 
    }
    \colnumbers
    \startdata
     % M-band
     & 0.3$\times$ & 4500 & 2400 & 1800 & 1300 \\
     \textit{M} & 1$\times$ & 4500 & 2400 & 1700 & 1300  \\
     & 3$\times$ & 4500 & 2200 & 1500 & 1100  \\ [0.3 cm]
    \enddata
\end{deluxetable}

\setlength{\tabcolsep}{5pt}
\begin{deluxetable}{cccccc}\label{tab:H2O}
    \tablecolumns{6}
    \tablewidth{1pt}
    \tablecaption{Minimum S/N$_{res}$ required for $\geq 5\sigma$ detection of H$_{2}$O.}
    \tablehead{\colhead{Observing} & \colhead{Insolation} & \multicolumn{4}{c}{Spectral Resolution} \\
    \cline{3-6}
    \colhead{Band} & \nocolhead{} & \colhead{25,000} & \colhead{50,000} & \colhead{75,000} & \colhead {100,000} 
    }
    \colnumbers
    \startdata
    % J-band
     & 0.3$\times$ & - & 2600 & 1900 & 1200 \\
    \textit{J} & 1$\times$ & - & 2800 & 2200 & 1300  \\
     & 3$\times$ & 3000 & 1600 & 1200 & 700  \\ [0.5 cm]
     % H-band
     & 0.3$\times$ & - & 4500 & 3000 & 2200  \\
     \textit{H} & 1$\times$ & - & 4000 & 2800 & 2000 \\
     & 3$\times$ & 3500 & 1800 & 1200 & 900 \\ [0.5 cm]
     % K-band
     & 0.3$\times$ & - & - & 4000 & 3000  \\
     \textit{K} & 1$\times$ & - & - & 4000 & 3000 \\
     & 3$\times$ & 5000 & 2800 & 1900 & 1500 \\ [0.5 cm]
     % L-band
     & 0.3$\times$ & - & 4500 & 3500 & 2600  \\
     \textit{L} & 1$\times$ & - & 4000 & 3000 & 2600  \\
     & 3$\times$ & 3000 & 1500 & 1300 &  1100 \\ [0.5 cm]
     % M-band
     & 0.3$\times$ & 5000 & 2800 & 1900 & 1300 \\
     \textit{M} & 1$\times$ & 4000 & 2400 & 1600 & 1200  \\
     & 3$\times$ & 1800 & 1000 & 700 & 500  \\ [0.3 cm]
    \enddata
\end{deluxetable}

\setlength{\tabcolsep}{5pt}
\begin{deluxetable}{cccccc}\label{tab:CH4}
    \tablecolumns{6}
    \tablewidth{1pt}
    \tablecaption{Minimum S/N$_{res}$ required for $\geq 5\sigma$ detection of CH$_{4}$. There are no $\geq 5\sigma$ detections with S/N$_{res}$ $\leq 5000$ for \textit{J}, \textit{H}, and \textit{M} bands.}
    \tablehead{\colhead{Observing} & \colhead{Insolation} & \multicolumn{4}{c}{Spectral Resolution} \\
    \cline{3-6}
    \colhead{Band} & \nocolhead{} & \colhead{25,000} & \colhead{50,000} & \colhead{75,000} & \colhead {100,000} 
    }
    \colnumbers
    \startdata
     % K-band
     & 0.3$\times$ & - & 3000 & 2000 & 1600  \\
     \textit{K} & 1$\times$ & - & - & - & - \\
     & 3$\times$ & - & - & - & - \\ [0.5 cm]
     % L-band
     & 0.3$\times$ & 1100 & 700 & 450 & 300  \\
     \textit{L} & 1$\times$ & - & - & - & -  \\
     & 3$\times$ & - & - & - &  - \\ [0.5 cm]
    \enddata
\end{deluxetable}

\subsection{1$\times$ GJ 1214b Insolation}
Three findings are consistently true regardless of wavelength range considered. First, as indicated in Table \ref{tab:Haze}, the haze is always the easiest opacity source to detect since it has the lowest required S/N$_{res}$ for detection, which is 500 or less in all cases considered. Thus, if one achieved the S/N$_{res}$ necessary to detect a molecule like CO in the atmosphere of one of these planets, one would necessarily have the required S/N$_{res}$ to rule out a completely clear atmosphere as well. Again, this is made possible through the s$_{g}$ and s$_{f}$ terms in the log-likelihood function and would thus be difficult if not impossible to detect using the classic CCF approach.

\indent Second, increasing the spectral resolution appears to have diminishing returns, i.e. increasing the spectral resolution from R $\sim$ 25,000 to 50,000 yields the greatest decrease in the required S/N$_{res}$ (often a factor of 2) but increasing the spectral resolution further does not yield quite as dramatic of a change in the required S/N level. This is because once enough strong spectral lines are resolved to clearly identify the presence of a molecule, adding additional weaker lines from further increased spectral resolution is an increasingly marginal help.  However, increasing to higher spectral resolution may still lead important gains in precision on parameters beyond detection of molecules, such as molecular abundances, temperature structure, or wind speeds that we do not consider in this work.

\indent Lastly, CH$_{4}$ is undetectable across all wavelength ranges and spectral resolutions for models with this insolation level. This inability to detect CH$_{4}$ is expected due to the lack of CH$_{4}$ above $\sim 10^{-4}$ bar (where these observations are most sensitive) as shown in Figure \ref{plot:AbundsProfs}.

\indent More specifically, here we break-down the band-by-band results:

\begin{figure*}[tp]
\caption{Detection significances for different opacity sources as function of spectral resolution and S/N$_{res}$ for \textit{M} band transmission spectra of the model with 1$\times$ GJ 1214b's insolation. All tested opacity sources except CH$_{4}$ are detectable in M band, though the haze and CO are detectable for a much wider range of combinations of S/N$_{res}$ and spectral resolution than CO$_{2}$ and H$_{2}$O.} 
\centering

\includegraphics[width=.49\textwidth]{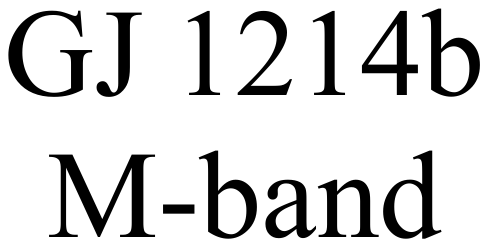}
\includegraphics[width=.49\textwidth]{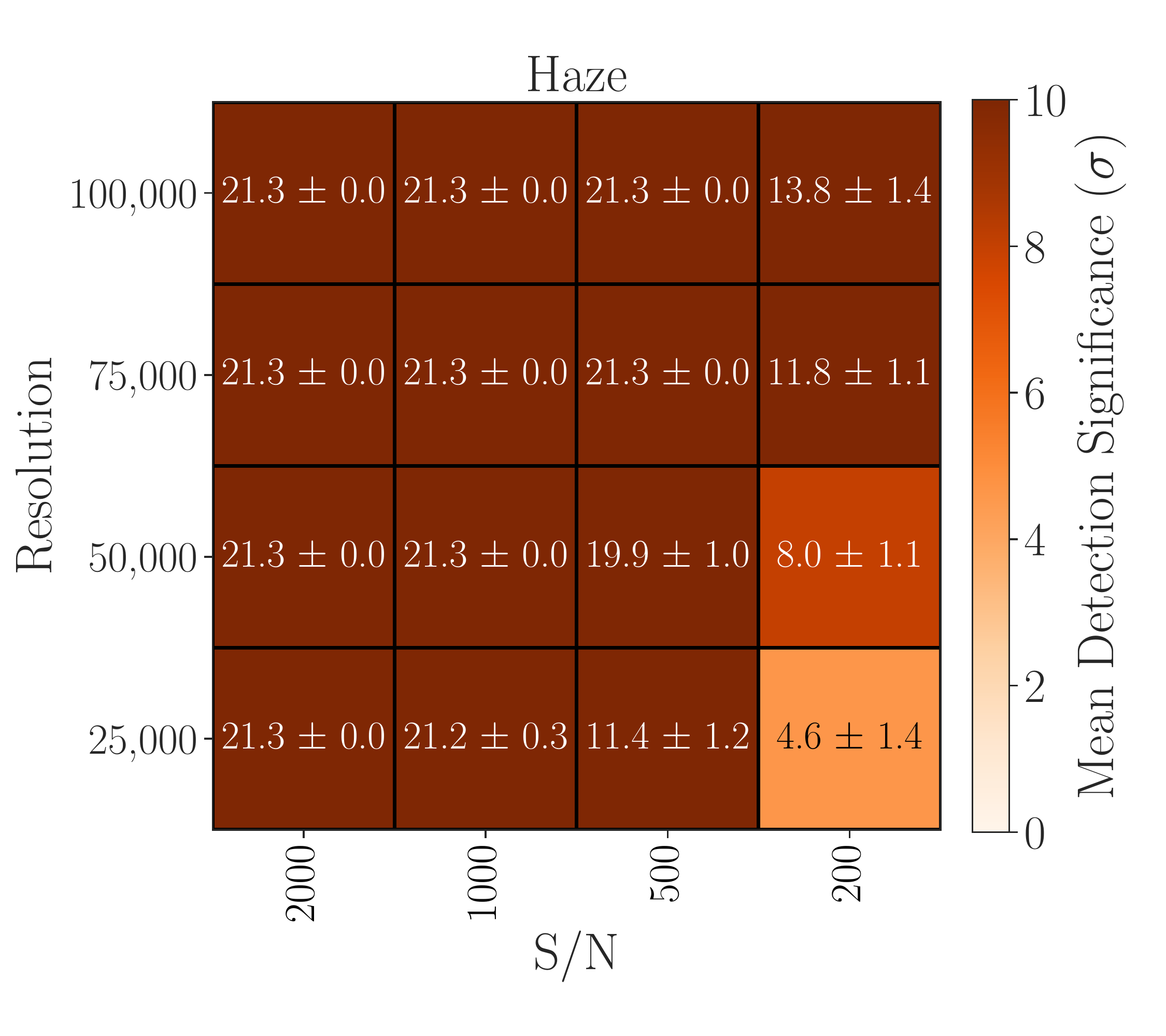}

\medskip

\includegraphics[width=.49\textwidth]{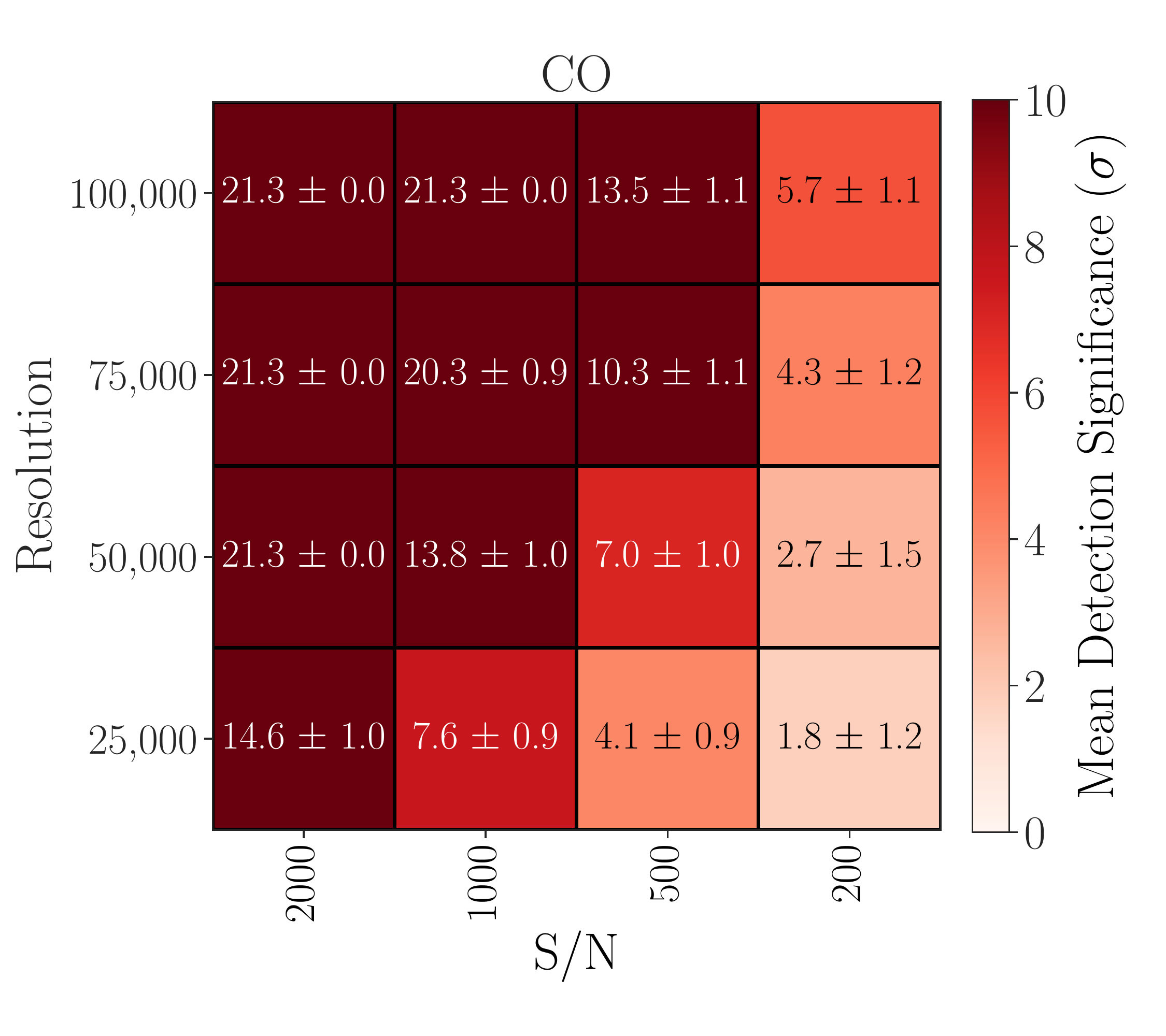}
\includegraphics[width=.49\textwidth]{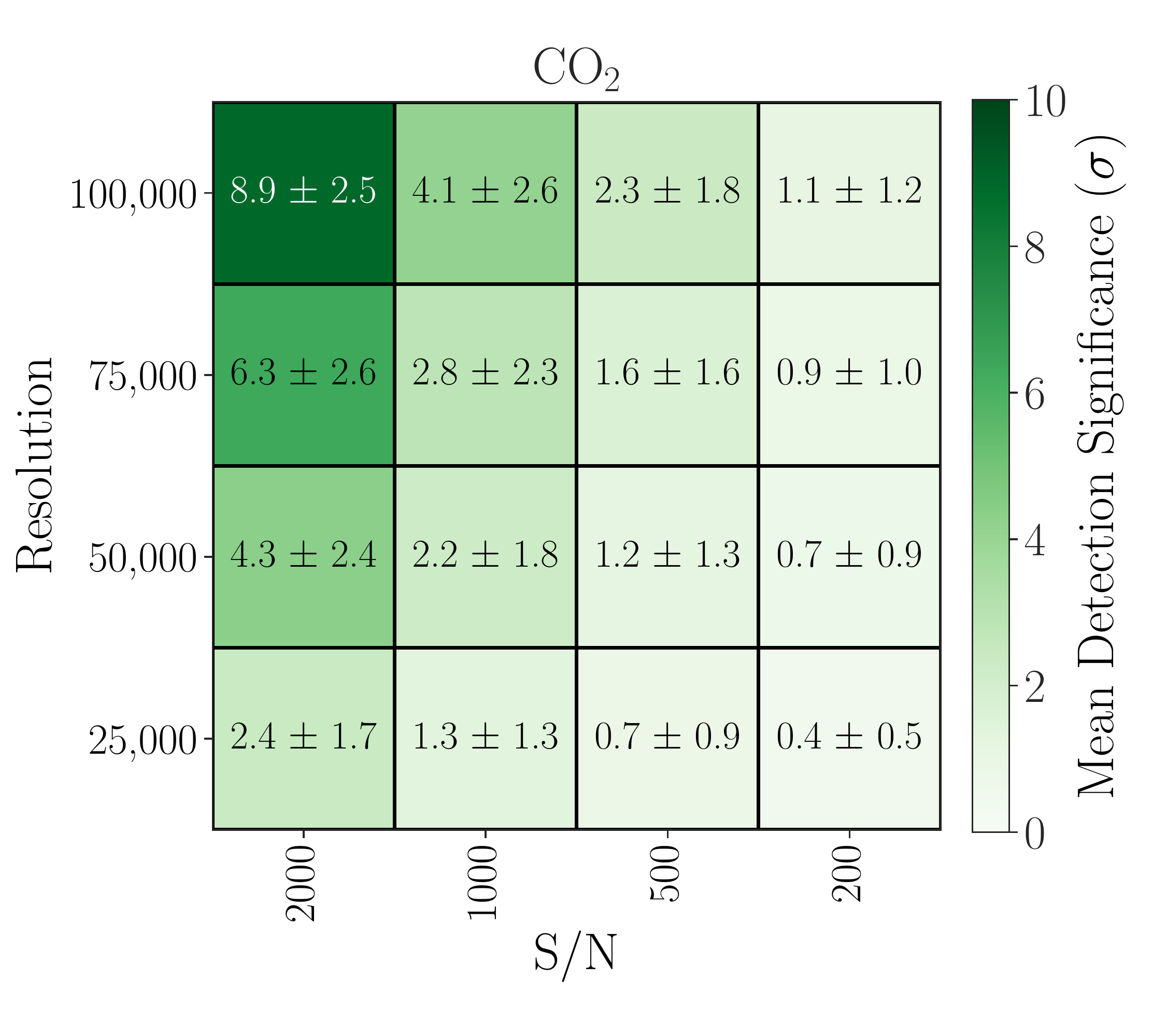}

\medskip

\includegraphics[width=.49\textwidth]{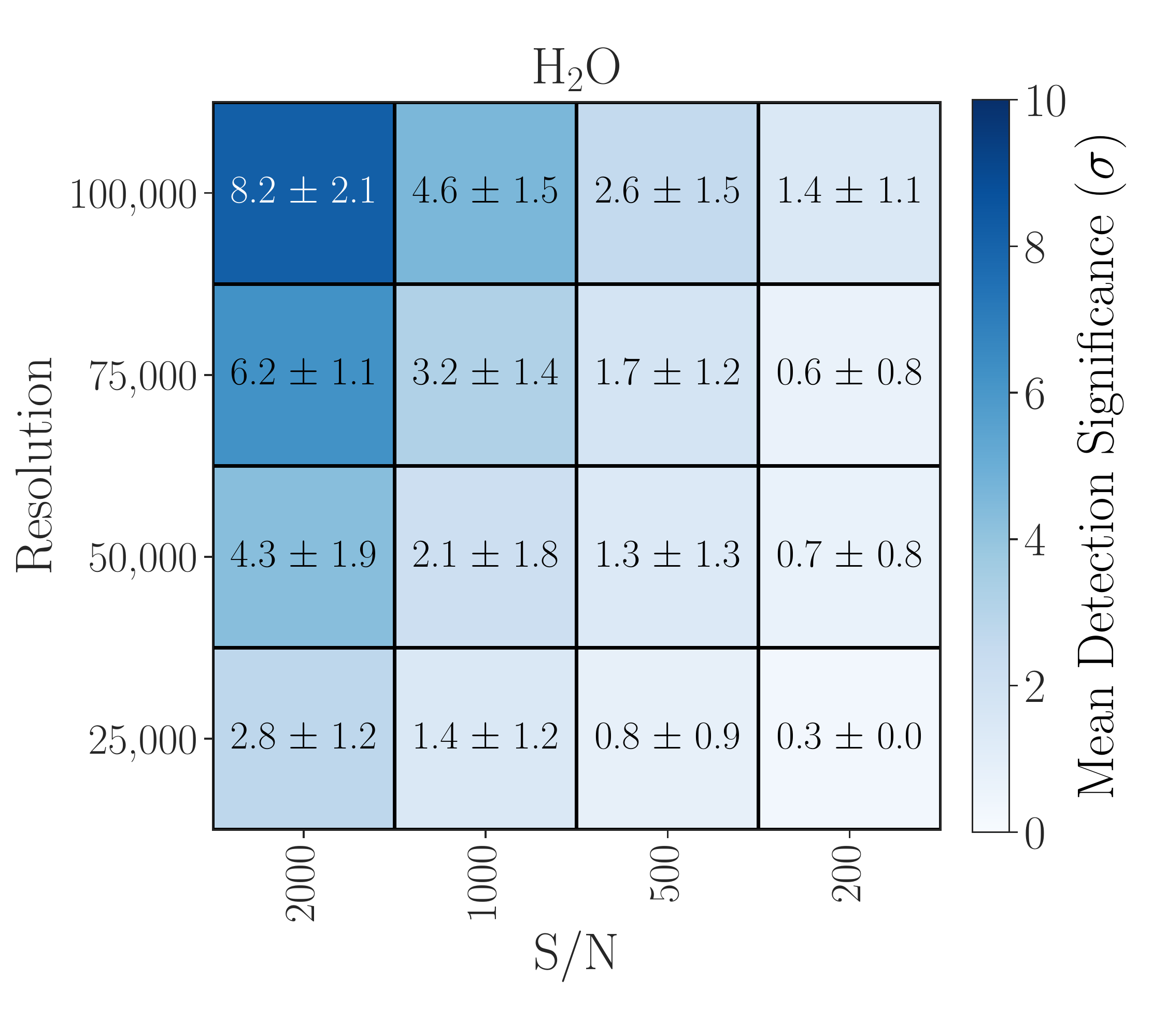}
\includegraphics[width=.49\textwidth]{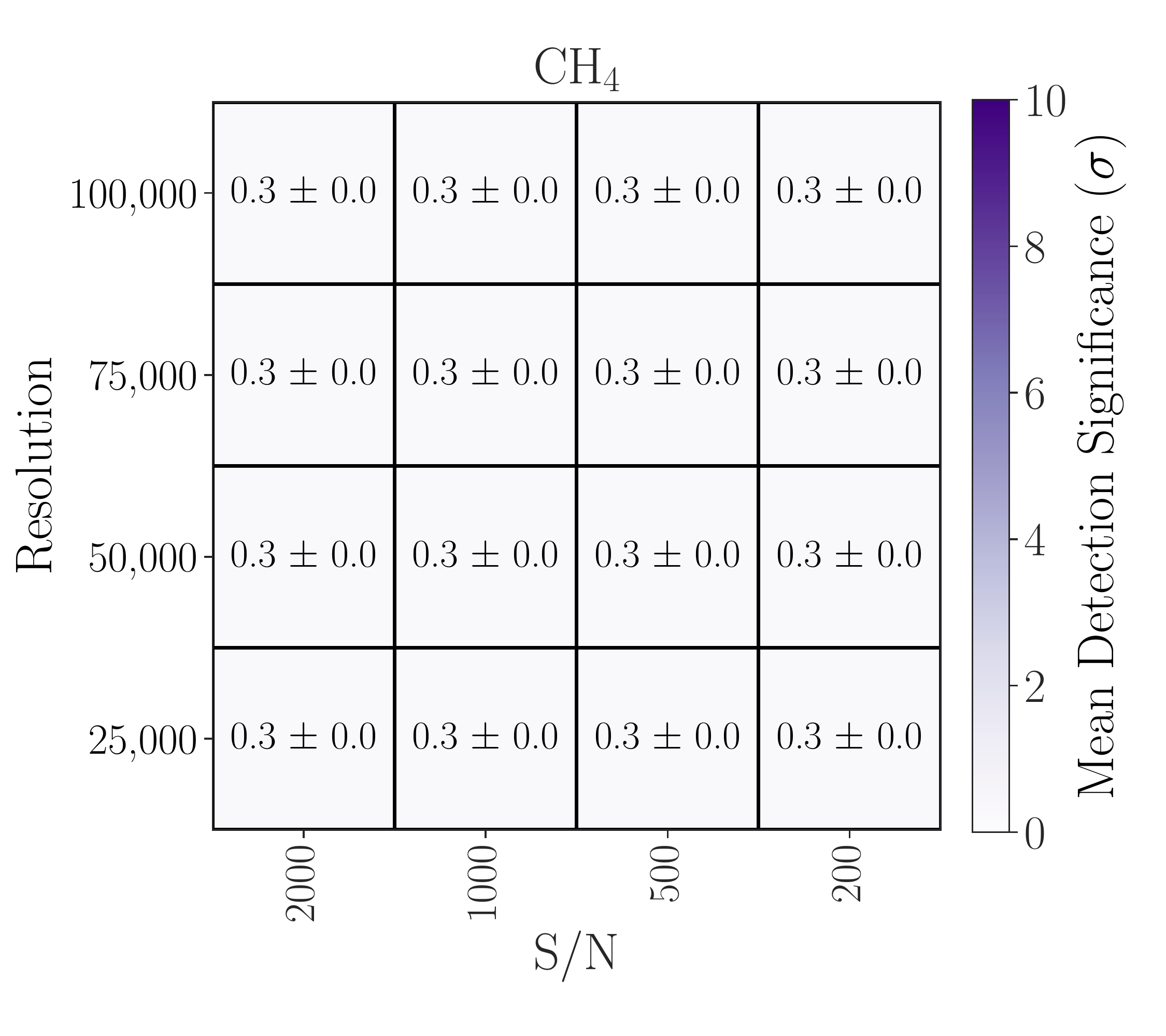}

\label{pics:Mband}
\end{figure*}

\begin{itemize}
    \item \textit{J} and \textit{H} Bands: H$_{2}$O is the only molecular opacity source detectable in addition to the haze. The required S/N$_{res}$ for detecting H$_{2}$O is lower in \textit{J} band than in \textit{H} - 2800 and 4000, respectively, for R $\sim$ 50,000.
    \item \textit{K} Band: CO is detectable for all spectral resolutions, while H$_{2}$O is detectable for R $\geq$ 75,000. However, H$_{2}$O requires much higher S/N spectra. For example, at a R $\sim$ 75,000, one needs an effective S/N$_{res}$ of 1100 to detect CO but 4000 for H$_{2}$O. 
    \item \textit{L} Band: H$_{2}$O is again the only detectable molecule. A higher S/N$_{res}$ is required to detect H$_{2}$O compared to \textit{J} or \textit{H}, with a required value of 4000 for R $\sim$ 50,000. 
    \item \textit{M} Band: CO, CO$_{2}$, and H$_{2}$O are all potentially detectable. Plots of the detection strength for all considered opacities as a function of selected spectral resolutions and S/Ns are shown in Figure \ref{pics:Mband}.  Detecting CO is significantly easier than in \textit{K}; at a R $\sim$ 75,000, one only needs an effective S/N$_{res}$ of 250 to detect CO. In addition, CO$_{2}$ is only detectable in this wavelength range, though it does require a higher S/N$_{res}$ than that needed to detect CO (for example, 1700 at a R $\sim$ 75,000). H$_{2}$O requires comparable S/N$_{res}$ to CO$_{2}$, but with spectra at R$\sim$100,000, one could detect CO, CO$_{2}$, and H$_{2}$O with S/N$_{res}$ $\geq$ 1100. Thus, the \textit{M} band is overall the most promising observing band for detecting 2 or more molecular opacity sources at once. However, the high thermal background in \textit{M} band may make these observations more challenging as discussed in Section \ref{subsec:observ}.
\end{itemize}

In general, the bands in which different molecules are detectable is our work here do reflect previous observational results. H$_{2}$O has been detected in \textit{J} band \citep{Alonso2019}, \textit{K} band \citep[e.g.][]{Hawker2018}, \textit{L} band \citep[e.g.][]{Birkby2013,Birkby2017,Piskorz2018}, and data covering 0.95 - 2.45 $\mu$m simultaneously \citep{Brogi2018,Guilluy2019}.  CO has been repeatedly detected in \textit{K} band \citep[e.g.][]{Snellen2010, Flagg2019}. To date, no molecules have been reported for high-resolution ground-based \textit{M} band spectra of an exoplanet atmosphere.

\cite{deKok2014} investigated the optimal wavelength ranges to detect different molecules in exoplanet atmospheres with the 2014-era CRIRES instrument on the VLT. Similar to our results presented above, they also find that CO is best detected in \textit{K} and \textit{M} bands, while H$_2$O is detectable in \textit{J} - \textit{M} bands. However, they find the region around 3.5 $\mu$m as optimal for detecting multiple species (CO$_{2}$, CH$_{4}$, and H$_{2}$O), in contrast to our preference for \textit{M} band. As we will discuss later in Section \ref{subsec:observ}, the thermal background in \textit{M} band can make it functionally difficult to reach the required S/N$_{res}$ with current instruments. In addition, they assume a truth spectrum with only one molecule at a time and cross-correlate with a model of just that species; they note that CH$_{4}$ or H$_{2}$O lines could have a shielding effect that make other molecules more difficult to detect. Thus, our inability to also detect CO$_{2}$ in L band can be attributed to the higher cross sections of the many lines of CH$_{4}$ and H$_{2}$O in this region as shown in Figure \ref{fig:xsecs}.

\subsection{ 0.3$\times$ and 3$\times$ Insolation}
\indent Since we would like to be able to observe a wider range of planets than just GJ 1214b, we also investigate how changing the stellar insolation effects our results, looking at models with 0.3$\times$ and 3$\times$ the true insolation of GJ 1214b. The primary effect is the change over of the dominant carbon bearing species with temperature: CH$_{4}$ dominating in the 0.3$\times$ case and CO/CO$_{2}$ in the 3$\times$ scenario, with an overlap/transition for the 1$\times$ case (Figure \ref{plot:AbundsProfs}).  

\subsubsection{0.3$\times$ Insolation}
As above, we list our results by observing band, but to simplify the discussion, only summarize the differences from the 1$\times$ insolation case. In some cases the haze opacity requires a slightly higher S/N$_{res}$ to detect than the nominal insolation case, but it still remains the easiest opacity source to detect across all wavelength bands and resolutions. Overall, the decrease in temperature for the models with 0.3$\times$ GJ 1214 b\textsc{\char13}s insolation makes CO, CO$_{2}$, and H$_{2}$O slightly harder to detect, while these are the only models where CH$_{4}$ is detectable. Though the volume mixing ratio of H$_{2}$O is actually slightly higher in the cooler case than that of the nominal insolation, this increase in difficulty may be attributed to the decrease in scale height with decreasing temperature which makes spectral features smaller. However, interference from the many CH$_{4}$ lines may also make H$_{2}$O harder to detect.  To test this idea, we took a 0.3$\times$ insolation model in \textit{L} band (which has the strongest CH$_{4}$ features) without CH$_{4}$ but including all of our other opacity sources as the "truth" model and computed the required S/N$_{res}$ to detect H$_{2}$O in this case. When CH$_{4}$ lines are not included, the required S/N$_{res}$ to detect H$_{2}$O decreases for all tested spectral resolutions, indicating that interference from CH$_{4}$ is indeed a source of the increased difficulty in detecting H$_{2}$O for these cooler models.

\begin{itemize}
    \item \textit{J} and \textit{H} Bands: The required S/N$_{res}$ for detecting H$_{2}$O is slightly lower in \textit{J} band and higher in \textit{H} band when compared to the 1$\times$ results. For example, with R$\sim$ 50,000 spectra, detecting H$_{2}$O would require S/N$_{res} \geq$ 2600 in \textit{J} band and S/N$_{res} \geq$ 4500 in \textit{H} band, compared to 2800 and 4000 for the 1$\times$ insolation case, respectively.
    \item \textit{K} Band: Detecting CO requires a much higher S/N$_{res}$ due to its decreased abundance, and is not detectable at R$\leq$25,000 for any of our noise scenarios. CH$_{4}$, which is not detectable in the nominal 1$\times$ models, is more readily detectable than CO and H2O in this cooler scenario (i.e. with S/N$_{res} \geq$2000 at R$\sim$75,000, compared to 3500 for CO and 4000 for H$_{2}$O). For the highest resolution case (R $\sim$ 100,000), one could detect CO, H$_{2}$O, and CH$_{4}$ with a S/N$_{res} \geq$ 3000. 
    \item \textit{L} Band: CH$_{4}$ and H$_{2}$O are both detectable. Due to the proximity to the v3 band, detecting CH$_{4}$ requires a much lower S/N$_{res}$ than in \textit{K}, with a lower limit of 300 in the R$\sim$100,000 case. A slightly higher S/N$_{res}$ is required to detect H$_{2}$O than in the 1$\times$ insolation case; S/N$_{res} \geq$ 3500 rather than 3000 for R$\sim$75,000.
    \item \textit{M} Band: In all cases the required S/N$_{res}$ to detect CO, CO$_{2}$, and H$_{2}$O is higher than for the nominal models. For R$\sim$ 100,000, CO, CO$_{2}$, and H$_{2}$O require S/N$_{res} \geq$ 400, 1300, and 1300, respectively (compared to 200, 1300, and 1200 for the 1$\times$ insolation case). 
\end{itemize}

\subsubsection{3$\times$ Insolation}
The hotter 3$\times$ insolation models show qualitatively similar detection behaviour to the 1$\times$ case. In most cases the haze opacity requires a slightly lower S/N$_{res}$ to detect than the nominal insolation case, but the change is often small. The biggest difference is that CO, CO$_{2}$ and H$_{2}$O are all easier to detect, sometimes by up to a factor of 2 decrease in the required S/N$_{res}$. 
\begin{itemize}
    \item \textit{J} and \textit{H} Bands: The required S/N$_{res}$ for detecting H$_{2}$O is approximately 2$\times$ lower than in the 1$\times$ insolation case for all resolutions in both bands, likely due to the increase in the size of the features in these bands compared to the nominal insolation case as seen in Figure \ref{fullspecs}.
    \item \textit{K} Band: Both CO and H$_{2}$O are detectable, but again require lower S/N$_{res}$ for a 5$\sigma$ detection. This decrease in required S/N$_{res}$ is $\sim$ 40\% for CO, while closer to $\sim$ 50\% for H$_{2}$O. Notably, in this hotter scenario, H$_{2}$O is detected at R$\leq$50,000 for all explored S/N cases, in contrast to the nominal 1$\times$ insolation case where it is only detectable at higher resolutions. 
    \item \textit{L} Band: The required S/N$_{res}$ for H$_{2}$O detection is again roughly 50\% lower than needed in the 1$\times$ insolation case across spectral resolutions. 
    \item \textit{M} Band: The required S/N$_{res}$ to detect CO, CO$_{2}$, and H$_{2}$O is lower than in the nominal insolation case, with the largest effect for H$_{2}$O. With spectra at R$\sim$100,000, one could detect CO, CO$_{2}$, and H$_{2}$O simultaneously with S/N$_{res}$ $\geq$ 1100 (compared to 1300 in the 1$\times$ insolation case). 
\end{itemize}

\subsection{Observing Multiple Bands}
\begin{figure*}
\centering 
\subfloat[\textit{J} Band]{%
  \includegraphics[width=0.45\textwidth]{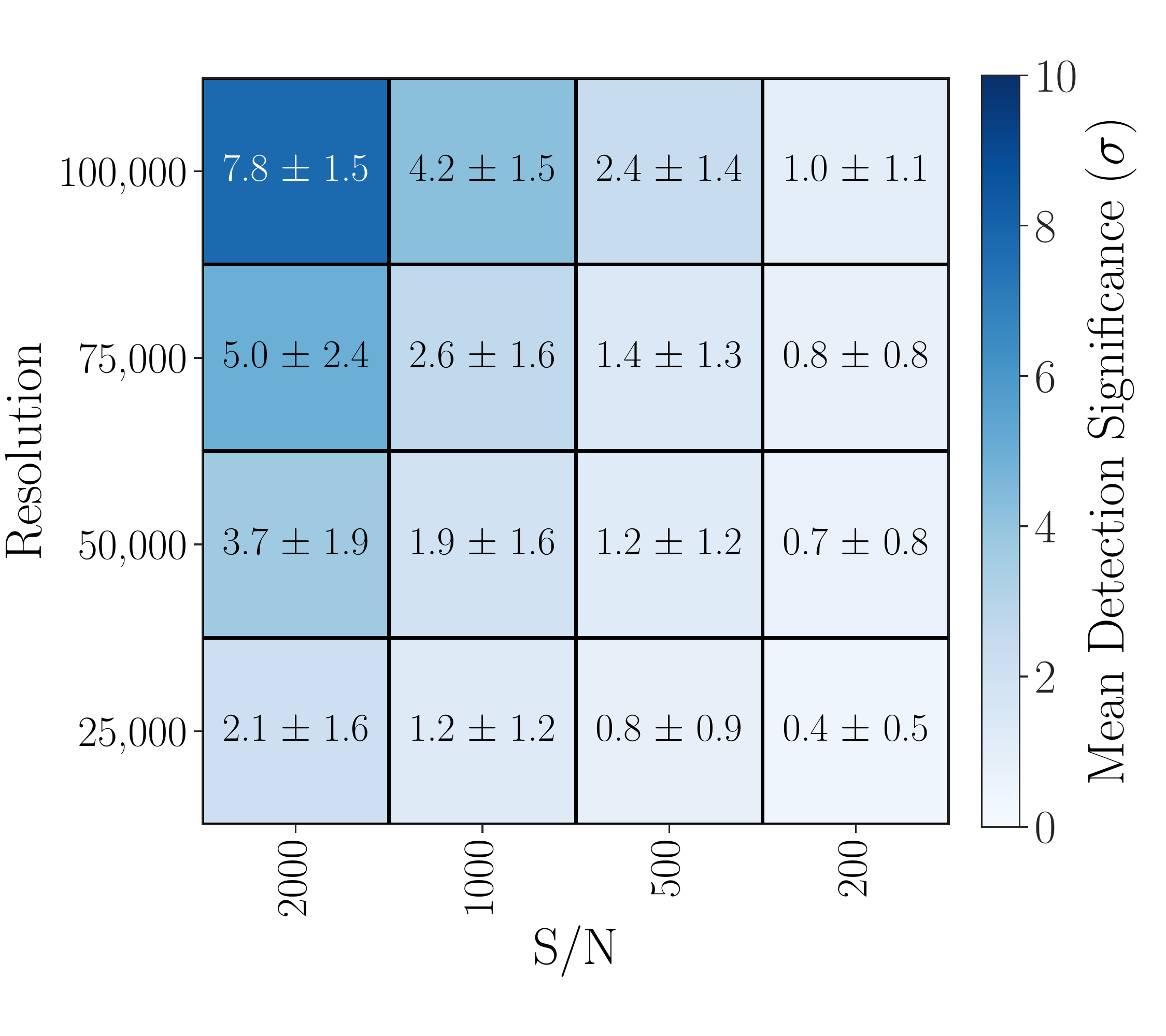}%
  \label{plot:Water_J}%
}
\subfloat[\textit{J} and \textit{H} Band]{%
  \includegraphics[width=0.45\textwidth]{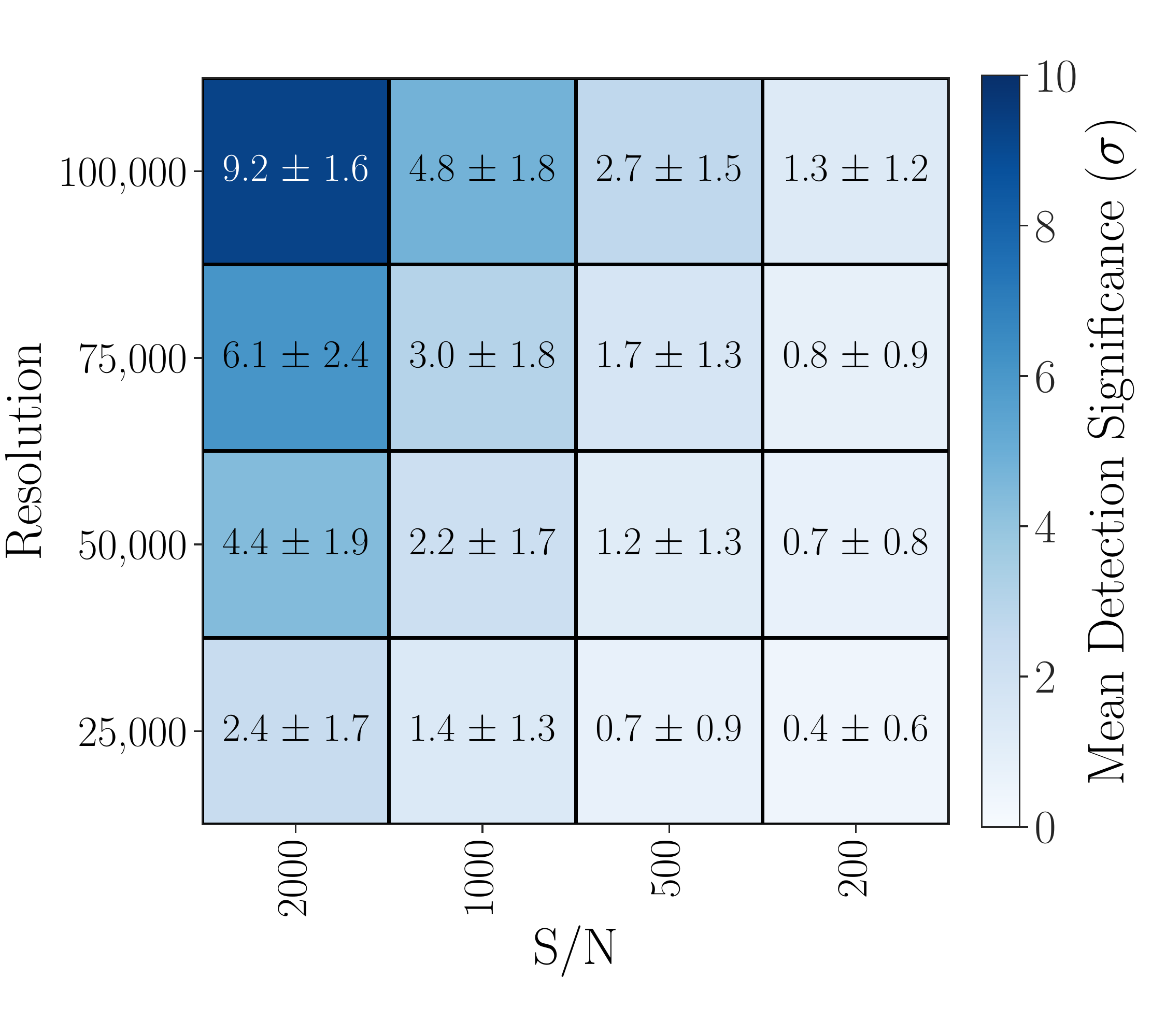}%
  \label{plot:Water_JH}%
}\qquad
\subfloat[\textit{K} Band]{%
  \includegraphics[width=0.45\textwidth]{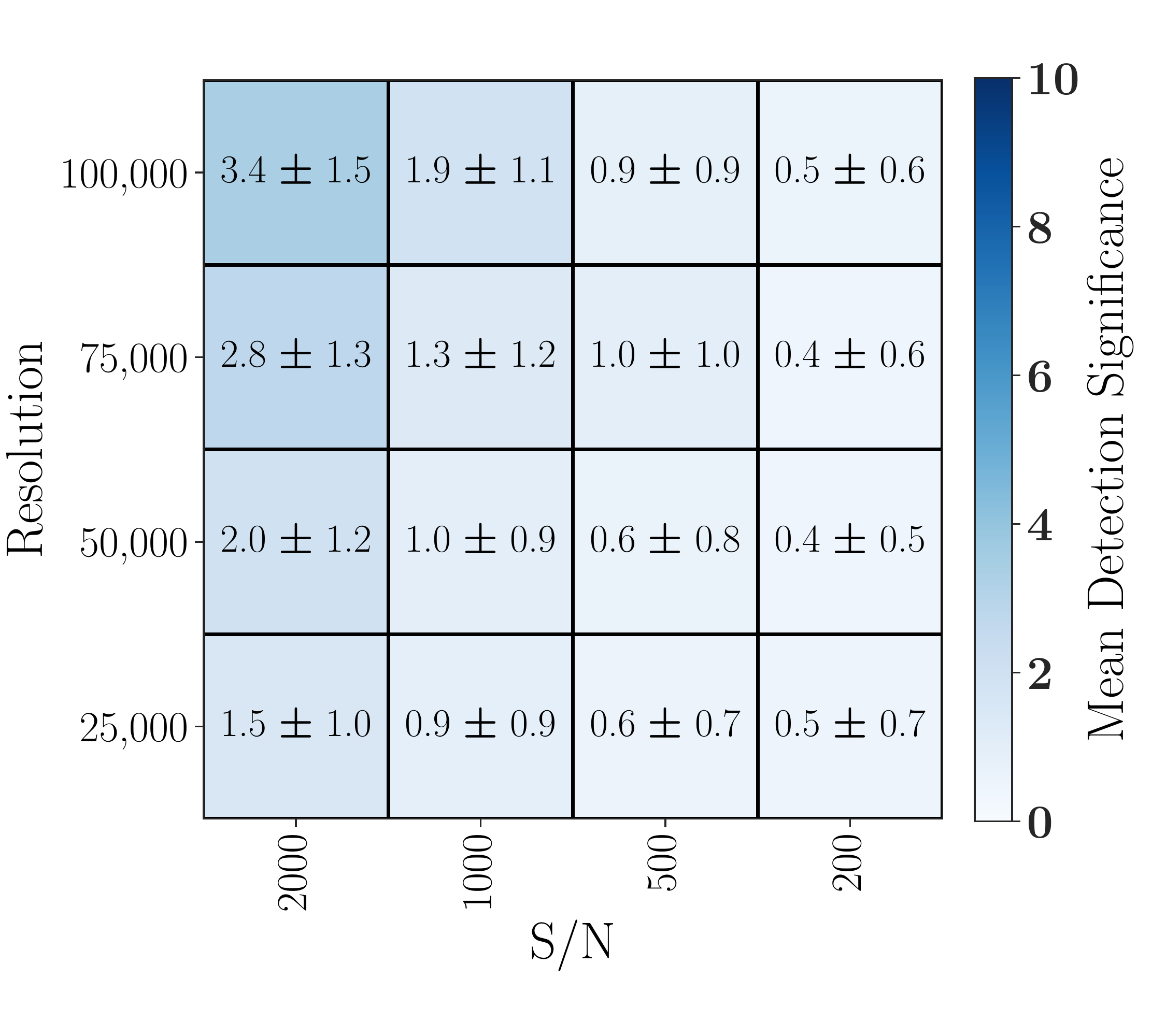}%
  \label{plot:Water_K}%
}
\subfloat[\textit{J}, \textit{H}, and \textit{K} Band]{%
  \includegraphics[width=0.45\textwidth]{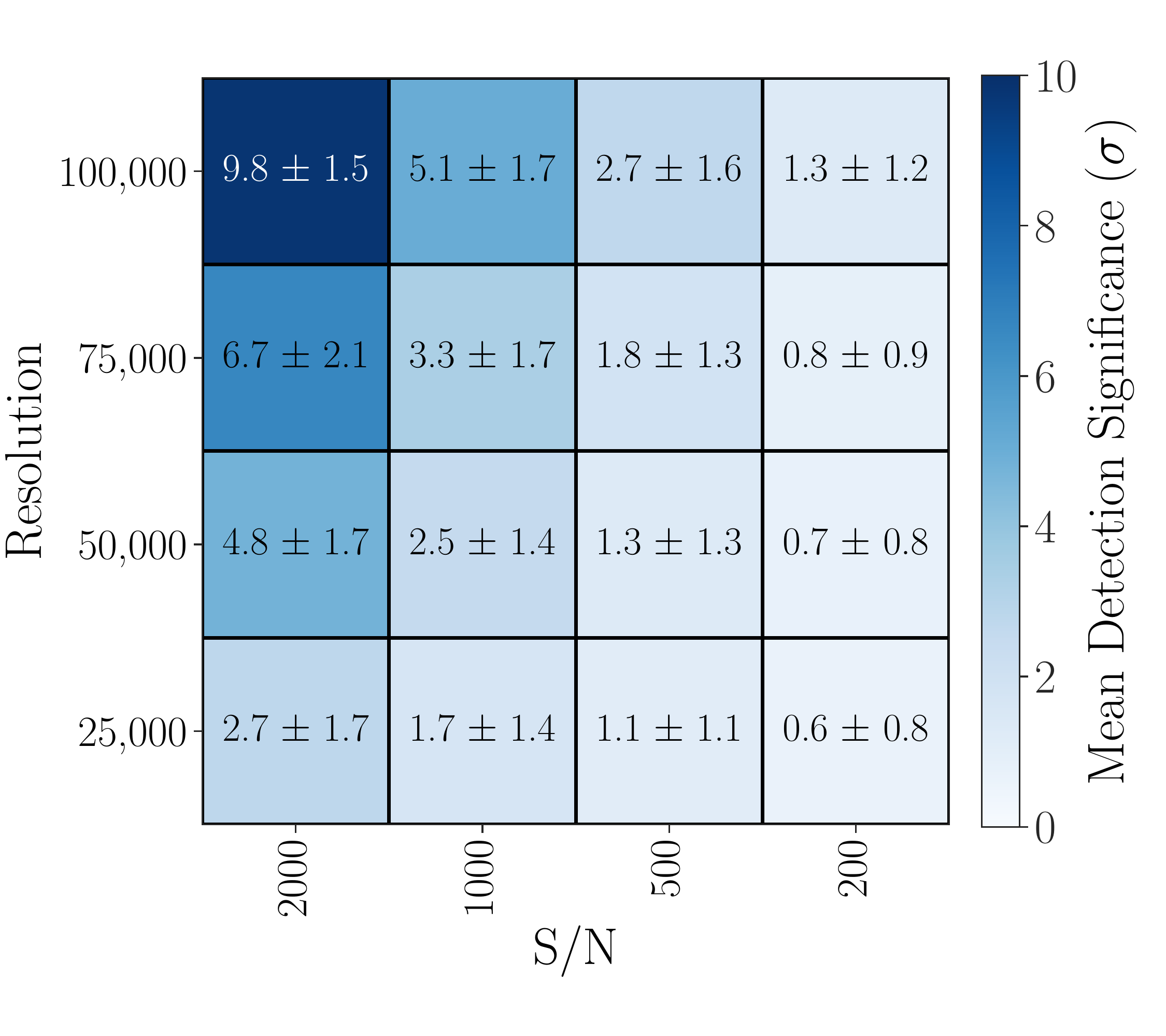}%
  \label{plot:Water_JHandK}%
}
\caption{Detection significances for H$_{2}$O for different observing bands as a function of spectral resolution and S/N per resolution element of the transmission spectrum. Observing \textit{J} and \textit{H} bands simultaneously increased the detection significance of H$_{2}$O as shown in \ref{plot:Water_J} and \ref{plot:Water_JH}. Similarly, while the H$_{2}$O detection in \textit{K} band is marginal in \ref{plot:Water_K}, adding \textit{J} and \textit{H} bands allows for a much stronger detection of H$_{2}$O.} 
\label{fig:H2OMultBands}
\end{figure*}

\indent Modern instruments are now able to observe multiple atmospheric windows simultaneously, often some selection of \textit{J}, \textit{H}, and \textit{K} bands, including CARMENES \citep{Quirrenbach2016}, NIRPS \citep{Wildi2017}, IGRINS \citep{Park2014}, GIANO \citep{Origlia2014}, and SPIRou \citep{Artigau2014}.  Access to multiple bands in one exposure allows for more spectral lines of a molecule to be observed, strengthening the signal of that molecule, as the molecular detection S/N scales as the $\sqrt{N_{lines}}$. Over these wavelength ranges, H$_{2}$O in particular has millions of spectral lines in each band. Figure \ref{fig:H2OMultBands} demonstrates the effect of increasing wavelength range on H$_{2}$O detection significance. Figure \ref{plot:Water_JH} shows a clear improvement when combining \textit{J} and \textit{H} bands over just observing \textit{J} band alone (Figure \ref{plot:Water_J}). Furthermore, H$_{2}$O is more difficult to observe in \textit{K} band, as shown in Figure \ref{plot:Water_K} and expected since H$_{2}$O has larger cross sections in \textit{J} and \textit{H} bands as shown in Figure \ref{fig:xsecs}. However, H$_{2}$O is much more detectable at fixed spectral resolution or S/N$_{res}$ when \textit{J} and \textit{H} band are also observed (Figure \ref{plot:Water_JHandK}). Thus, instruments that can observe \textit{J}-\textit{K} band simultaneously will more readily detect H$_{2}$O and CO(which is detectable in \textit{K} band but neither of the other two bands; see Table \ref{tab:CO} with the same observations.

\indent Looking to future instrumentation, GMTNIRS \citep{Jaffe2016} is a proposed high-resoltuion spectrograph for the GMT that would cover 1 to 5 $\mu$m in one exposure. Table \ref{tab:AllBands} shows the minimum S/N$_{res}$ required to detect each opacity source when observing all bands (\textit{J}-\textit{M}) simultaneously. While the required S/N$_{res}$ to detect H$_{2}$O is substantially less than when looking at any single band as shown in Table \ref{tab:H2O}, other opacity sources are not as benefited by simultaneous wavelength coverage, due to their narrower span or dominance by H$_{2}$O over most bands. In particular, the minimum S/N$_{res}$ required to detect CO and CO$_{2}$ is in most cases identical to that required for when considering solely \textit{M} band spectra, so adding in other bands does not make either molecule easier to detect. Similarly, the minimum S/N$_{res}$ reported for CH$_{4}$ in Table \ref{tab:AllBands} are identical to those for analyzing solely \textit{L} band spectra from Table \ref{tab:CH4}. Thus, H$_{2}$O is the opacity source, under these specific atmospheric conditions, most benefited by an instrument with wide instantaneous wavelength coverage. In scenarios in which other broad-band absorbing molecules dominate, say, higher metallicity where CO/CO$_{2}$ are more prominent, or much cooler where CH$_{4}$ dominates the full near infrared, the influence of multiple bands on specific molecular detections would undoubtedly change. We leave this detailed analysis to a future study.

\setlength{\tabcolsep}{5pt}
\begin{deluxetable}{cccccc}\label{tab:AllBands}
    \tablecolumns{6}
    \tablewidth{0pt}
    \tablecaption{Minimum S/N$_{res}$ required for $\geq 5\sigma$ detection of each opacity source for transmission spectra that cover observing bands \textit{J}-\textit{M}} simultaneously.
    \tablehead{\colhead{Opacity} & \colhead{Insolation} & \multicolumn{4}{c}{Spectral Resolution} \\
    \cline{3-6}
    \colhead{Source} & \nocolhead{} & \colhead{25,000} & \colhead{50,000} & \colhead{75,000} & \colhead {100,000} 
    }
    \colnumbers
    \startdata
    % CO
     & 0.3$\times$ & 1400 & 800 & 600 & 400 \\
    CO & 1$\times$ & 700 & 350 & 250 & 200  \\
     & 3$\times$ & 400 & 250 & 150 & 150  \\ [0.5 cm]
     % CO2
     & 0.3$\times$ & 4500 & 2400 & 1800 & 1200  \\
     CO$_{2}$ & 1$\times$ & 4500 & 2400 & 1700 & 1300 \\
     & 3$\times$ & 4500 & 2200 & 1500 & 1100 \\ [0.5 cm]
     % H2O
     & 0.3$\times$ & 2800 & 1500 & 1100 & 800  \\
     H$_{2}$O & 1$\times$ & 2600 & 1400 & 1100 & 800 \\
     & 3$\times$ & 1200 & 700 & 500 & 350 \\ [0.5 cm]
     % CH4
     & 0.3$\times$ & 1100 & 700 & 450 & 300  \\
     CH$_{4}$ & 1$\times$ & - & - & - & -  \\
     & 3$\times$ & - & - & - &  - \\ [0.5 cm]
     % Haze
     & 0.3$\times$ & 150 & 100 & 100 & 100  \\
     Haze & 1$\times$ & 150 & 100 & 100 & 50 \\
     & 3$\times$ & 100 & 100 & 50 & 50 \\ [0.5 cm]
    \enddata
\end{deluxetable}

\subsection{Photochemical Products}
\indent Though our results so far have only included four molecules (CO, CO$_{2}$, H$_{2}$O, and CH$_{4}$), other molecules could be important opacity sources for these planets in the near infrared. In particular, photochemical products could affect the high-resolution transmission spectra at these wavelengths and their detection could provide an avenue to distinguish between a photochemical haze and equilibrium condensate clouds. However, not all potential molecules have high fidelity line lists at the temperatures and pressures necessary for generating these models. \cite{Hawker2018} and \cite{Cabot2019} have both presented evidence of HCN in the atmospheres of hot Jupiters using high-resolution spectroscopy, indicating the possibility of detecting this molecule with current line lists. Thus, to test whether photochemical products would be detectable with these kinds of observations, we focus on HCN as an illustrative example using the line list from \cite{Harris2008}. 

\indent As in Section \ref{subsec:trspec}, we create a new ``truth" spectrum that includes HCN in addition to our other opacity sources. We use the results of the photochemical model cited in Section \ref{subsec:trspec} to determine the HCN abundance. We can then use the log(L) method described in Section \ref{sec:methods} to compare to the models without HCN and quantify our ability to detect the molecule as a function of S/N$_{res}$ and spectral resolution.  We find that \textit{L} band is the only observing band where HCN is detectable; so far the only spectra used to detect HCN in \cite{Hawker2018} and \cite{Cabot2019} covered 3.18 - 3.27 $\mu m$. The minimum S/N$_{res}$ needed to detect HCN is listed in Table \ref{tab:HCNdet}. HCN is only detectable for the hottest models and highest spectral resolutions. However, as shown in Figure \ref{fig:xsecs}, HCN has its strongest features between 3 and 3.2 $\mu$m. If spectra starting at 3.1 $\mu$m instead of 3.15 $\mu$m can be obtained, HCN could be much easier to detect - for example, the required S/N$_{res}$ for 3$\times$ insolation and R $\sim$ 100,000 decreases from 3500 to 1000.

\indent We also explore how unaccounted for photochemical products could affect the detectability of the major molecular species. Our ``truth" spectrum inlcudes our standard set of opacities (CO, CO$_{2}$, H$_{2}$O, CH$_{4}$, and H$_{2}$/He CIA) plus HCN, but compare to models without HCN (e.g., an ``incorrect" model).  We find that the detection S/N values do not change in any case,  even for \textit{L} band where HCN is detectable (Table \ref{tab:HiddenHCN}). This consistency suggests that our above results are robust against missing absorbers.  However, since HCN is not easily detectable, the minimum S/N$_{res}$ for a detection may be slightly higher than reported here if there are unaccounted-for molecules that would more significantly affect the high-resolution transmission spectrum. 

\setlength{\tabcolsep}{5pt}
\begin{deluxetable}{ccccc}\label{tab:HCNdet}
    \tablecolumns{5}
    \tablewidth{1pt}
    \tablecaption{Minimum S/N$_{res}$ required for $\geq 5\sigma$ detection of HCN with \textit{L} band spectra.}
    \tablehead{\colhead{Insolation} & \multicolumn{4}{c}{Spectral Resolution} \\
    \cline{2-5}
    \nocolhead{} & \colhead{25,000} & \colhead{50,000} & \colhead{75,000} & \colhead {100,000} 
    }
    \colnumbers
    \startdata
    0.3$\times$ & - & - & - & - \\
    1$\times$ & - & - & - & -  \\
    3$\times$ & - & - & 4000 & 3500  \\ [0.5 cm]
    \enddata
\end{deluxetable}

\setlength{\tabcolsep}{5pt}
\begin{deluxetable}{cccccc}\label{tab:HiddenHCN}
    \tablecolumns{6}
    \tablewidth{0pt}
    \tablecaption{Minimum S/N$_{res}$ required for $\geq 5\sigma$ detection of each opacity source for transmission spectra in \textit{L} band when HCN is included in the observed spectrum but not the comparison models. CO and CO$_{2}$ are not observable with S/N$_{res}$ $\leq 5000$.}
    \tablehead{\colhead{Opacity} & \colhead{Insolation} & \multicolumn{4}{c}{Spectral Resolution} \\
    \cline{3-6}
    \colhead{Source} & \nocolhead{} & \colhead{25,000} & \colhead{50,000} & \colhead{75,000} & \colhead {100,000} 
    }
    \colnumbers
    \startdata
     % H2O
     & 0.3$\times$ & - & 4500 & 3500 & 2600  \\
     H$_{2}$O & 1$\times$ & - & 4000 & 3000 & 2600 \\
     & 3$\times$ & 3000 & 1500 & 1300 & 1100 \\ [0.5 cm]
     % CH4
     & 0.3$\times$ & 1100 & 700 & 450 & 300  \\
     CH$_{4}$ & 1$\times$ & - & - & - & -  \\
     & 3$\times$ & - & - & - &  - \\ [0.5 cm]
     % Haze
     & 0.3$\times$ & 300 & 200 & 150 & 100  \\
     Haze & 1$\times$ & 400 & 250 & 200 & 150 \\
     & 3$\times$ & 300 & 200 & 100 & 100 \\ [0.5 cm]
    \enddata
\end{deluxetable}

\section{Discussion} \label{sec:disc}
\subsection{Observability with Current and Future Instrumentation}\label{subsec:observ}
In Section 3 we presented the required S/N per resolution element to detect molecules over a range of stellar insolation levels ($T_{\rm eff}$ = 412, 557, and 733 K) and observational parameters (R $\sim$ 25,000 - 100,000, 50 $\leq$ S/N$_{res}$ $\leq$ 5000, and \textit{J} - \textit{M} bands). However, the exposure times required to reach these S/N$_{res}$ will vary depending on host star brightness, sky background, telescope aperture size, and instrument sensitivity. While a detailed instrument/observational investigation is outside the scope of this work, we can focus on GJ 1214b as an example of how these factors might affect observations. 

We used a simplified noise model for estimating instrumental S/N$_{res}$ as a function of exposure time to determine the observability of molecules as a function of source brightness, telescope size, resolving power, and wavelength. The estimated measured S/N per resolution element depends upon the number of photons received from the source (which itself depends on exposure time, throughput, collecting area, brightness, and resolution), thermal background (which depends on emissivity, throughput, etendue A$\Omega$, and temperature), and instrumental quantities such as total throughput ($\tau \sim$~ 0.05), emissivity ($\epsilon = 0.3$), and noise properties of the detector (dark current, read noise), which all combine into:

\begin{equation}\label{eq:SNR}
    S/N = \frac{S \times T}{\sqrt{S \times T + N_{exp} \times (T_{exp} \times (BKGD + DC) + RN^2)}}
\end{equation}

where $S$ is the signal, $T$ is the total exposure time, $N_{\rm exp}$ is the number of exposures, $T_{\rm exp}$ is the time for a single exposure (assumed to be 600 seconds), \emph{BKGD} is thermal background, \emph{DC} is the dark current, and \emph{RN} is read noise. 
\begin{figure*}
\centering 
\subfloat[]{%
  \includegraphics[width=0.45\textwidth]{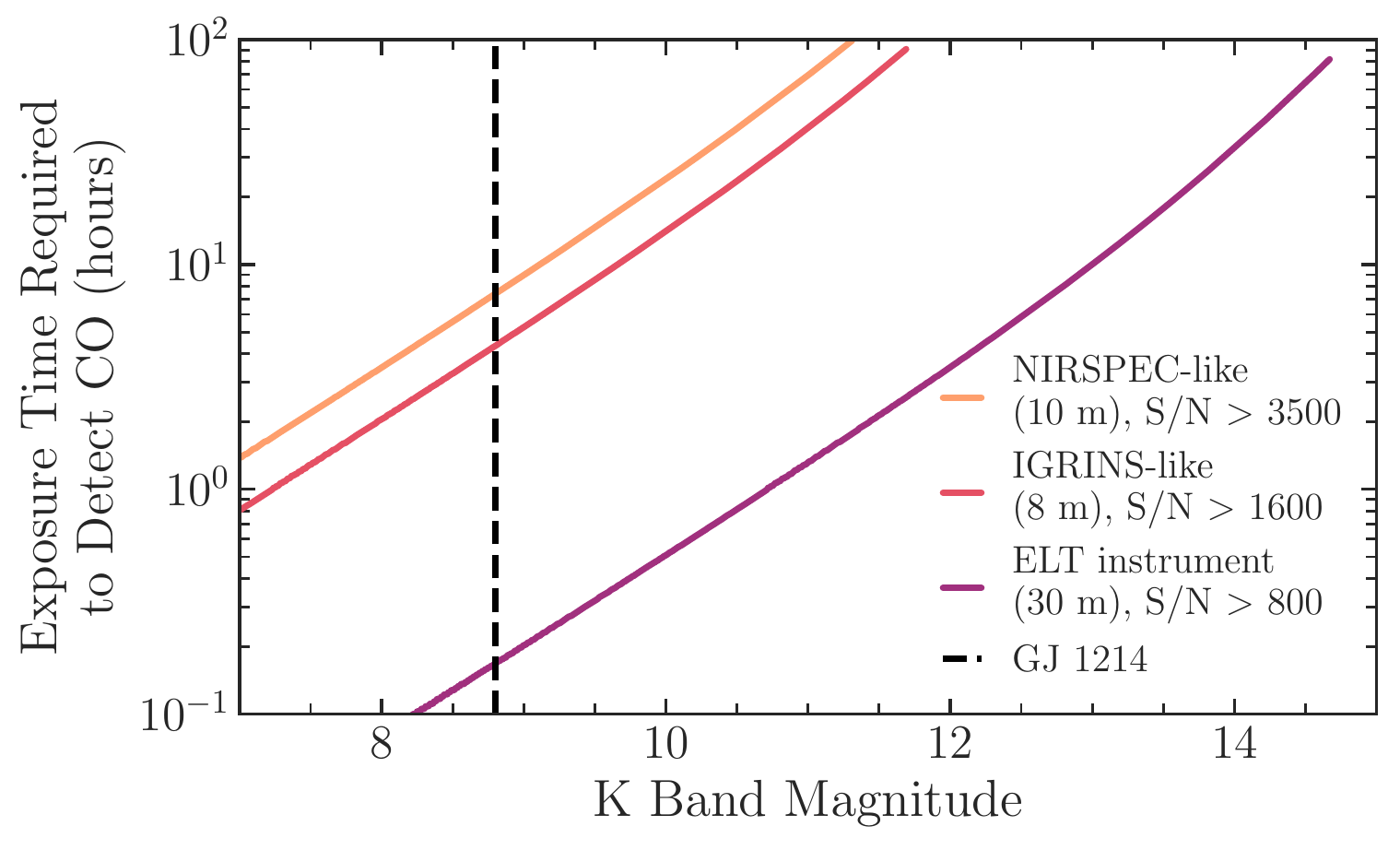}%
  \label{plot:Itime_CO_K}%
}\qquad
\subfloat[]{%
  \includegraphics[width=0.45\textwidth]{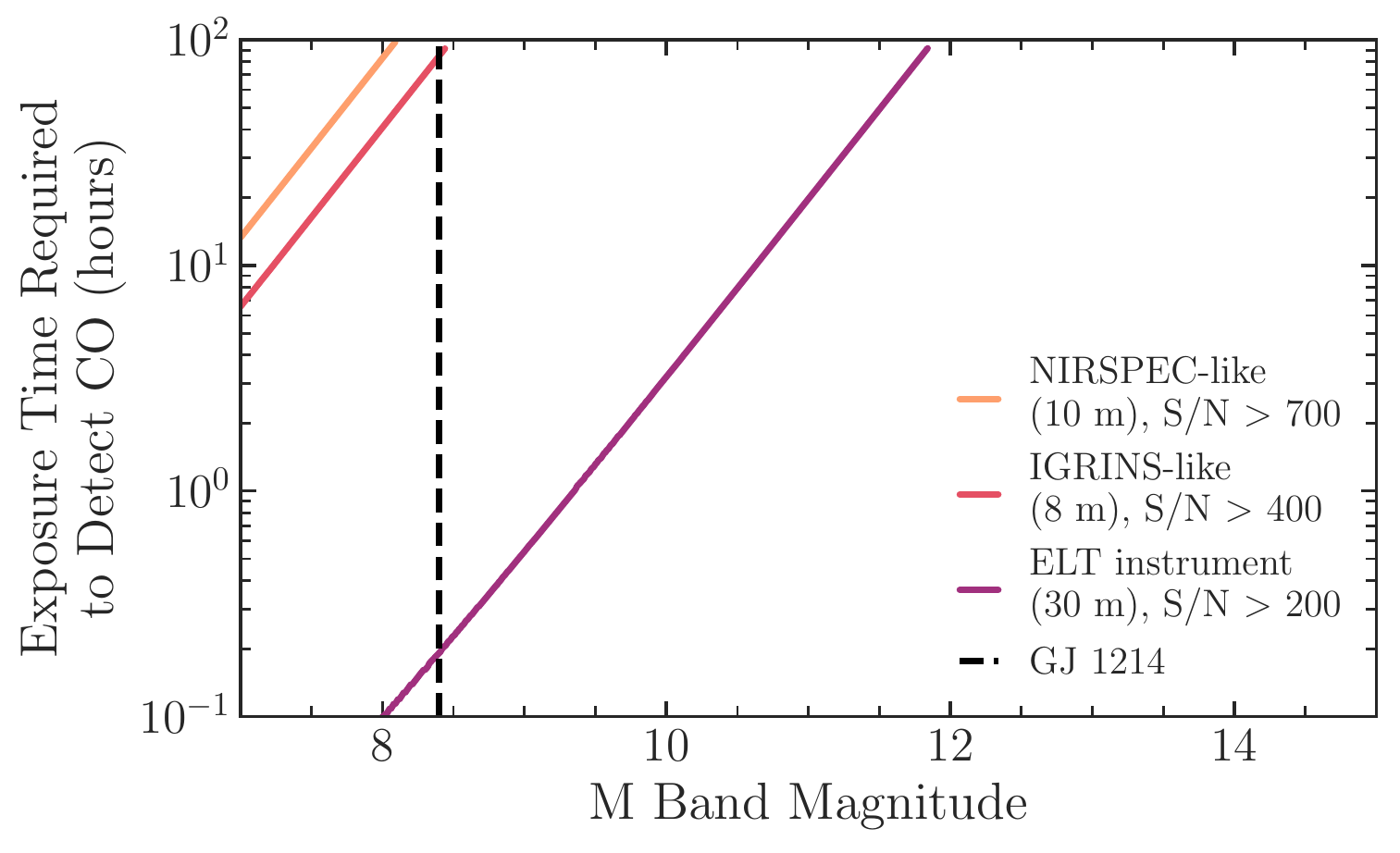}%
  \label{plot:Itime_CO_M}%
}
\caption{Estimated exposure time required to detect CO as a function of host star magnitude for different instruments in \textit{K} band (Figure \ref{plot:Itime_CO_K}) and \textit{M} band (Figure \ref{plot:Itime_CO_M}). The magnitude of GJ 1214 is marked by the dashed line. We consider three different types of instruments- a “NIRSPEC-like” R$\sim$25,000 spectrograph on a 10-m telescope, an “IGRINS-like” R$\sim$50,000 spectrograph on an 8-m telescope, and a proposed R$\sim$100,000 spectrograph behind AO on a 30-m telescope like the TMT. We find that although detecting CO requires a lower S/N$_{res}$ in \textit{M} band than \textit{K} band, \textit{K}-band observations are actually much more feasible, particularly for current instruments. However, a potential high-resolution spectrograph on a 30-m telescope would be able to detect CO in GJ 1214b in \textit{M} band with one transit.}
\label{fig:Itimes}
\end{figure*}

\begin{figure*}
\centering 
\subfloat[]{%
  \includegraphics[width=0.45\textwidth]{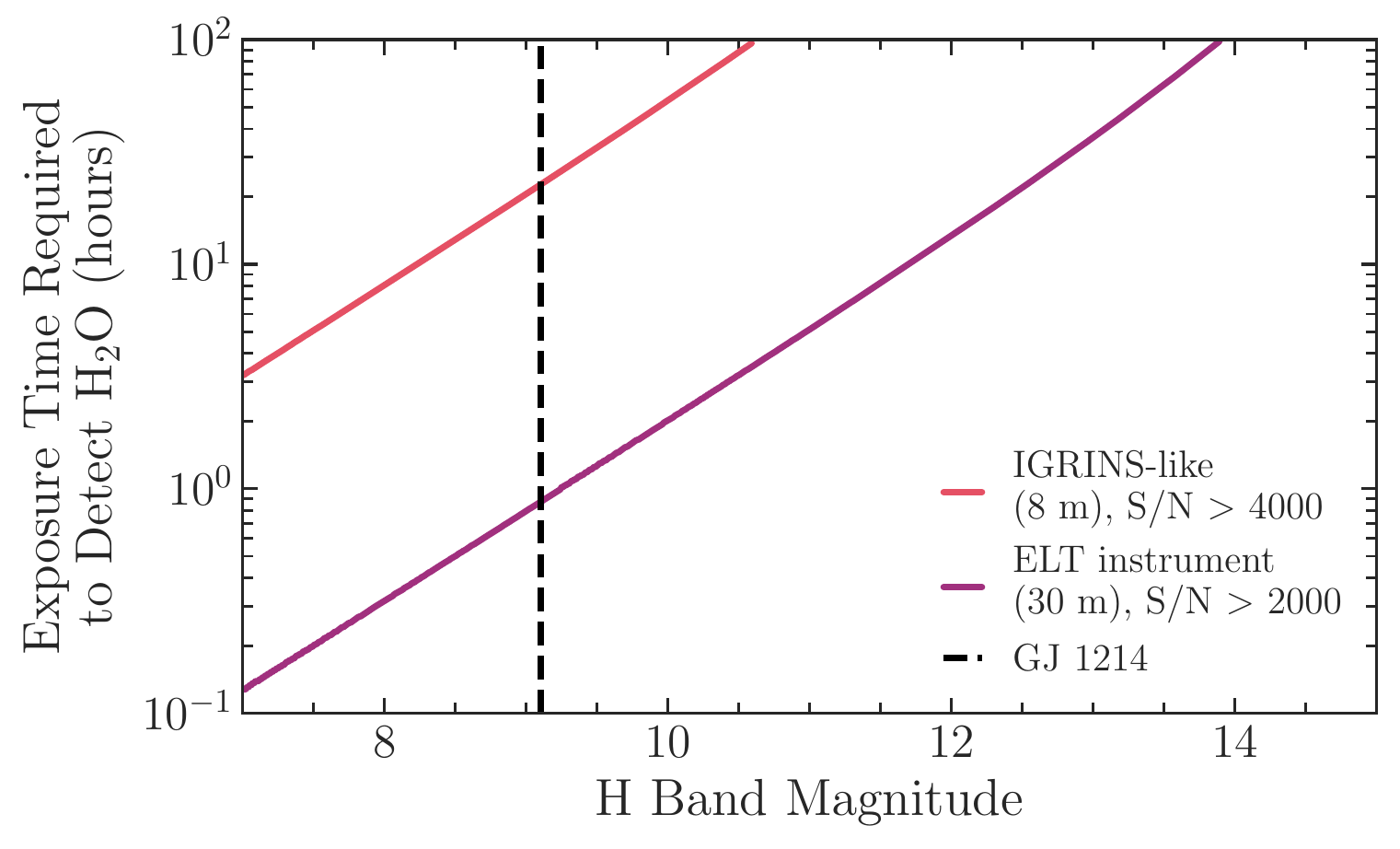}%
  \label{plot:Itime_H2O_H}%
}\qquad
\subfloat[]{%
  \includegraphics[width=0.45\textwidth]{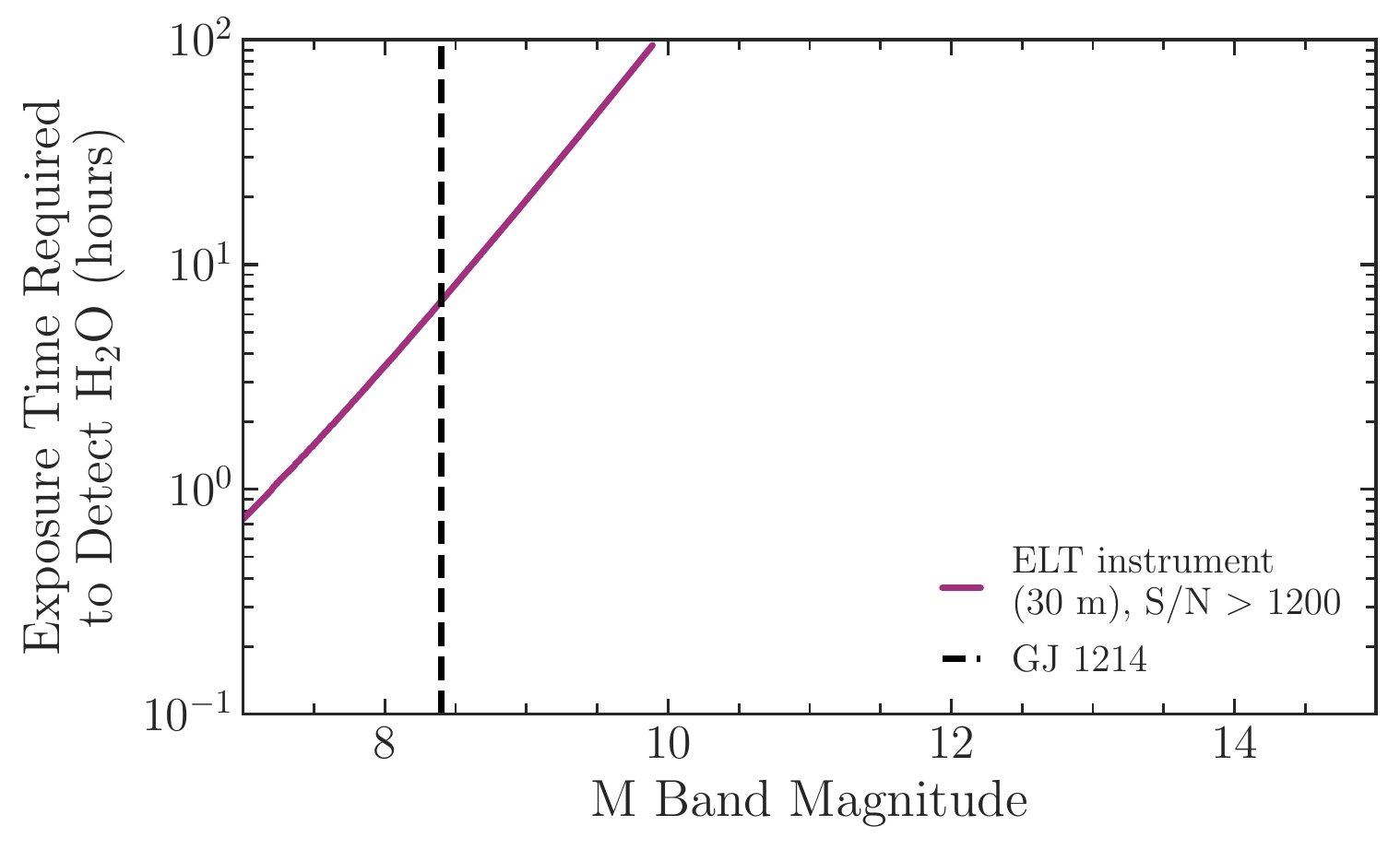}%
  \label{plot:Itime_H2O_M}%
}
\caption{Estimated exposure time required to detect H$_{2}$O as a function of host star magnitude for different instruments in \textit{H} band (Figure \ref{plot:Itime_H2O_H}) and \textit{M} band (Figure \ref{plot:Itime_H2O_M}). The magnitude of GJ 1214 is marked by the dashed line.  In \textit{H} band we did not have any H$_{2}$O detections with S/N$_{res} \leq$ 5000 with R $\sim$ 25,000, so in Figure \ref{plot:Itime_H2O_H} we just consider an “IGRINS-like” R$\sim$50,000 spectrograph on an 8-m telescope and a proposed R$\sim$100,000 spectrograph behind AO on a 30-m telescope like the TMT. Similar to Figure \ref{fig:Itimes}, we find that although detecting H$_{2}$O requires a lower S/N$_{res}$ in \textit{M} band than \textit{H} band, \textit{H}-band observations are more feasible, particularly for current instruments. However, a potential high-resolution spectrograph on a 30-m telescope would be able to detect H$_{2}$O in GJ 1214b in \textit{H} band with $\sim$ 7 hours of integration time.}
\label{fig:Itimes_H2O}
\end{figure*}

\begin{figure}[htbp]
\centering 
\includegraphics[width=.95\linewidth]{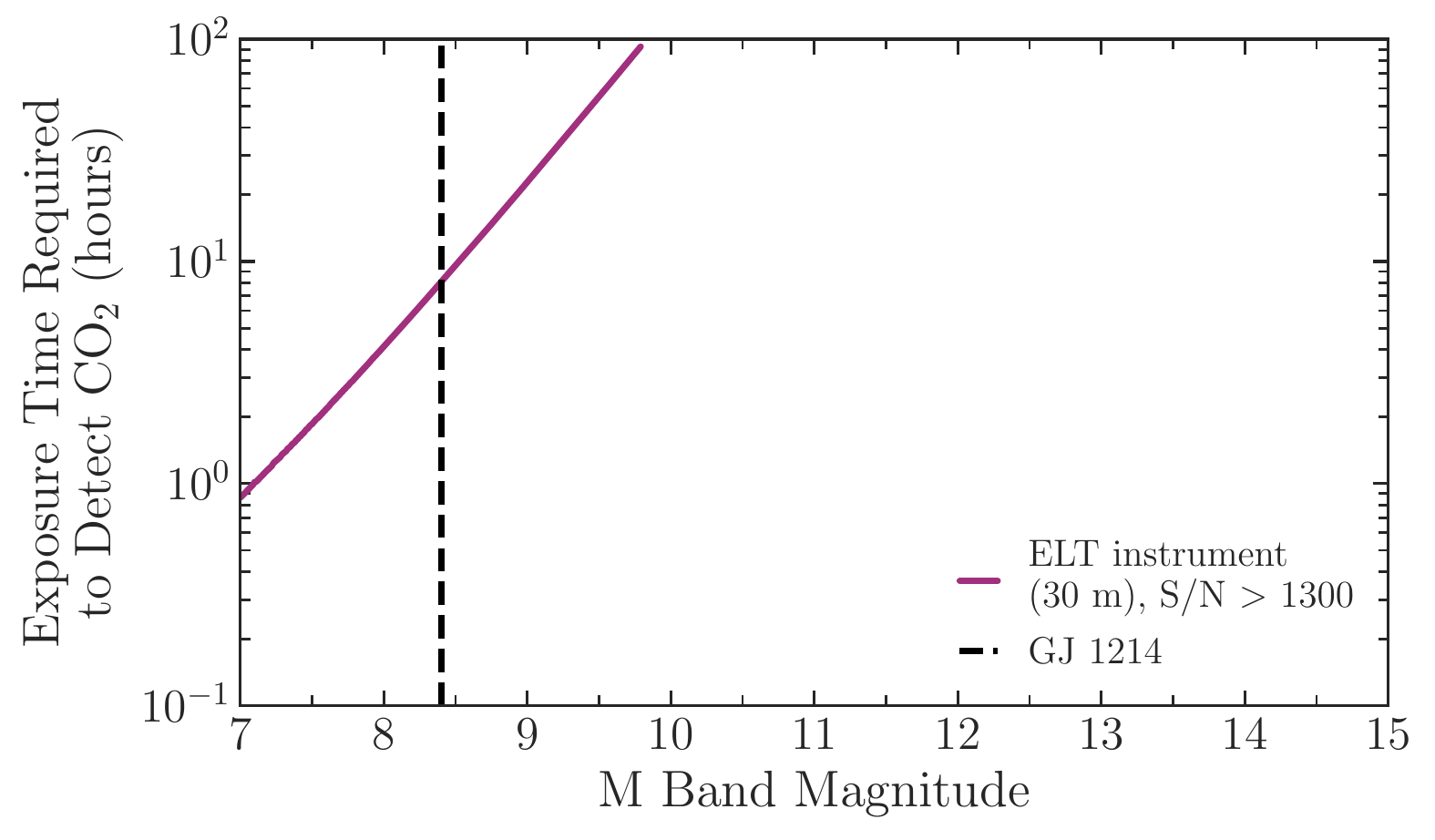}
\caption{Estimated exposure time required to detect CO$_{2}$ as a function of host star magnitude for different instruments in  \textit{M} band. The magnitude of GJ 1214 is marked by the dashed line.  Of our three investigated instrument types, only a proposed R$\sim$100,000 spectrograph behind AO on a 30-m telescope like the TMT can detect CO$_{2}$ in a reasonable amount of observing time ($\sim$ 8 hours for GJ 1214b).}
\label{fig:Itimes_CO2}
\end{figure}

Using this equation, it is then possible to estimate the S/N$_{res}$ expected for a given source magnitude, resolving power, and exposure time. For each molecule, we determined the minimum exposure time required to detect the molecule in the 1$\times$ insolation scenario (using GJ 1214b-like planet/star parameters) for varying source brightnesses. We simulated the exposure time vs. source brightness relation for three different scenarios: 1) $R\sim25,000$ instrument on a 10-m telescope, seeing-limited (e.g. NIRSPEC on Keck, \citealt{McLean1998}, \citealt{Martin2018}); 2) $R\sim50,000$ instrument on an 8-m telescope, seeing-limited (e.g. IGRINS on Gemini, \citealt{Park2014}, \citealt{Mace2018}); 3) $R\sim100,000$ instrument on a 30-m telescope, behind Adaptive Optics (AO), such as the proposed METIS on ELT \citep{Brandl2018} or MODHIS \citep{Mawet2019} on TMT. The dark current and read noise for the IGRINS-like and ELT instruments were assumed to be typical Teledyne values\footnote{\url{http://www.teledyne-si.com/products/Documents/H2RG\%20Brochure\%20-\%20September\%202017.pdf}} ; the NIRSPEC-like curve used measured values from that instrument\footnote{\url{https://www2.keck.hawaii.edu/inst/nirspec/Specifications.html}}. To easily normalize for each type of instrument, we assumed a ``seeing disk" of 20 pixels per resolution element (assuming the resolution element in the imaging and dispersion directions was 5 pixels by 4 pixels). We assume exposures with the maximum practical length one would want to take in a particular band due to sky lines and thermal background: 600 seconds for \textit{K} band and 30 seconds for \textit{M} band. We do not assume any overhead time that might occur, for example due to detector readout, which would increase the required amount of telescope time for these observations.  Each resolving power requires a different minimum S/N$_{res}$ to detect a molecule. For example, in \textit{K} band, to detect CO at 5$\sigma$ significance for 1$\times$ insolation, an R$\sim$25,000 instrument requires S/N$_{res}>$ 3500, an R$\sim$50,000 instrument needs S/N$_{res}>$ 1600, and an R$\sim$100,000 instrument needs S/N$_{res}>$ 800. 

Figure \ref{fig:Itimes} shows the results of our noise observability analysis. For \textit{K} band, shown in Figure \ref{plot:Itime_CO_K}, an instrument similar to NIRSPEC would take about 8 hours while the IGRINS-like instrument would take about 4 hours to detect CO. The  mean transit duration of GJ 1214b is $52.73^{+0.49}_{-0.35}$ minutes, which translates to $\sim$8 transits with NIRSPEC and $\sim$4 transits with IGRINS. An ELT instrument behind AO with R$\sim$100,000 would reduce this required time to around 6 minutes, so observing one full transit will be more than enough time. In contrast, for \textit{M} band observations, both current instruments would take greater than 100 hours to detect CO due to the much higher background noise in this wavelength range. An ELT instrument behind AO would take about 8 minutes to detect CO in this wavelength range as shown in Figure \ref{plot:Itime_CO_M}, so again one full transit would be more than enough time.  However, we note that these observing times assume good observing conditions and could readily double in the case of low seeing or slit losses. 

We performed similar observability calculations for H$_{2}$O and CO$_{2}$. For H$_{2}$O, we looked at \textit{H} and \textit{M} bands. For \textit{H} band spectra, we did not detect H$_{2}$O with S/N$_{res} \leq$ 5000 for R$\sim$25,000 (see Table \ref{tab:H2O}) so Figure \ref{plot:Itime_H2O_H} only shows the exposure times for the IGRINS-like and TMT instruments. \textit{H} band spectra with the IGRINS- like instrument would about 12 hours of total integration for H$_{2}$O to be detected. Detecting H$_{2}$O in \textit{M} band spectra, however, would take an inordinate amount of observing time ($>$ 100 hours) with current instruments but again should be more easily detectable with about 7 hours on a 30-m telescope with AO for GJ 1214b. Similarly, only the R$\sim$100,000 instrument on a 30-m telescope can detect CO$_{2}$ within any reasonable amount of time, $\sim$ 8 hours for GJ 1214b as seen in Figure \ref{fig:Itimes_CO2}.  

\subsection{Additional Caveats}

There are other caveats to consider when using the S/N$_{res}$ values in Tables 2-9 to plan observations for GJ 1214b or any other hazy sub-Neptune. First, when computing these values, we assumed all spectra covered the full wavelength range stated in Table 1, regardless of resolution, as the exact wavelength coverage varies between instruments. However, increasing spectral resolution may come with decreased wavelength coverage, leading to fewer spectral lines in the data, which would make these molecules more difficult to detect. 

In addition, GJ 1214b has a relatively slow radial velocity change over the course of one transit (12 km s$^{-1}$; \citealt{Crossfield2011}), compared to the hot Jupiters for which this technique has successfully been applied. Thus, it may be difficult to remove the quasi-stationary contamination from our atmosphere and the star while preserving the planet spectrum. 

Finally, we are assuming no contamination from stellar lines in our spectra (see Section 2.1). This is a good approximation for host stars with minimal spectral features in the observed wavelengths. However, for planets around M-dwarfs like GJ 1214, a myriad of stellar lines may make proper removal of the telluric and stellar contamination challenging. \cite{Brogi2016} and \cite{Schwarz2016} successfully modelled and removed stellar lines before removing the tellurics, but it remains to be seen if this technique can be successfully applied to M-dwarf spectra as more complex stellar models and data analysis are needed to remove overlapping stellar-planet molecular features \citep[e.g.,][]{Chiavassa2019}. Stellar variability may also pose a challenge, particularly for combining data from multiple nights, especially for M-dwarfs which are known to have high levels of magnetic activity \citep[e.g.,][]{Newton2016}.  As a result, a conservative approach may be to focus on detection and characterization of species  unlikely to be abundantly present in M-dwarf photospheres (e.g., CH$_{4}$, NH$_{3}$, HCN, etc.).

\section{Conclusions}\label{sec:conc}
In this work, we have investigated the feasibility of detecting molecules in the atmospheres of hazy sub-Neptunes with ground-based, high-resolution spectroscopy. To do so, we generated high-resolution transmission spectra of GJ 1214b analogs with a photochemical haze that matches the featureless low resolution transmission spectrum.  We considered two different metrics from the literature to quantify our detection significances: the cross correlation function (CCF) and a log likelihood function (log(L)) derived by \cite{Brogi2019}. While both metrics produced similar detection significances for molecules, only the log(L) was sensitive to the presence of the hazy opacity due to the additional terms that track the spectral variance relative to the data variance. Thus, we used the log(L) for the remainder of this work. However, our method relies on measuring the change in log(L) as we remove one opacity source at a time, indicating more care may need to be taken when determining which opacity sources to include in one's model.

We have calculated the minimum signal-to-noise (S/N) required for a $>5\sigma$ detection of each opacity source (CO, CO$_{2}$, H$_{2}$O, CH$_{4}$, and the haze) as a function of stellar insolation, spectral resolution, and wavelength range. Our key results are as follows.
\begin{enumerate}
    \item High resolution infrared transmission spectrum observations for hazy GJ 1214b analogs probe pressures around 1\,$\mu$bar and numerous molecular features can be detected for spectra that otherwise appear ``featureless" at $R\sim$100-1000.
    \item The haze is always the easiest opacity source to detect and observable in all observing bands. Thus, achieving the S/N per resolution element required to detect a molecule will also allow one to rule out a completely clear atmosphere.
    \item H$_{2}$O is detectable with S/N$_{res} \leq 5000$ for almost all combinations of spectral resolution and wavelength coverage, and is the only molecule observable in \textit{J} and \textit{H} bands, but always requires a higher S/N$_{res}$ than any other molecules observable in that band. In contrast, CO is only observable in \textit{K} and \textit{M} bands, but is the easiest molecule to detect. CO$_{2}$ is only observable in \textit{M} band; in fact, \textit{M} band is the observing band that requires the lowest S/N$_{res}$ to detect two or more molecules at once (although in practice requires long observing time due to thermal background noise).
    \item In general, increasing the stellar insolation of the model lowers the required S/N$_{res}$ for molecular detections. However, CH$_{4}$ is only detectable in the \textit{L} band spectra of the coldest models.
    \item HCN is detectable with \textit{L} band spectra for high resolution spectra of the hottest models. Detecting HCN, along with other potential photochemical products, could be a way to distinguish between equilibrium cloud and photochemical haze opacity in a planet's atmosphere.
\end{enumerate}

To further investigate the observability of these ultra-hazy sub-Neptunes, we used a simple model in Section \ref{sec:disc} to determine the feasibility of observing high-resolution transmission spectroscopy of planets with current and future instruments. We found that detecting CO and H$_{2}$O for GJ 1214b with current instruments observing in \textit{K} and \textit{H} bands, respectively, requires on the order of 10 hours of observing time. Furthermore, though a lower S/N$_{res}$ is required to detect these molecules in \textit{M} band, current instruments would need an unreasonable investment of observing time due to the high background. However, such an observation would be trivial behind AO on an ELT class telescope. In addition, a high resolution spectrograph on an actively cooled space telescope could observe in \textit{M} band without issues from thermal background.

As discussed in Section \ref{sec:disc}, more detailed, instrument-specific simulations for particular targets may be needed for careful observation planning. Such a study for ELT instrument concepts could help inform design decisions to maximize the information we can learn about these objects. Furthermore, an analogous study could be conducted of thermal emission spectra, which will likely require the S/N achievable with ELT instruments.  A number of sub-Neptunes will be observable with \emph{JWST}, albeit with lower spectral resolution. Analysis of simulated joint \emph{JWST} and ground-based high-resolution observations could pinpoint optimal observing strategies to take full advantage of these complementary data sets. As more sub-Neptunes around bright, nearby stars are discovered \citep[as expected from NASA's TESS mission;][]{Barclay2018}, considering the best application of observational techniques is essential to maximizing the scientific return on this abundant class of planets. 

\section*{Acknowledgments}
We are grateful to Dr. Gregory Mace for his assistance and advice on the instrumental observability section. Posting of this manuscript on the arXiv was coordinated with S. Ghandi et al. ECM is supported by an NSF Astronomy and Astrophysics Postdoctoral Fellowship under award AST-1801978.  MRL and JJF acknowledge the support of NASA XRP grant 80NSSC19K0293. JLB acknowledges funding from the European Research Council (ERC) under the European Union’s Horizon 2020 research and innovation program under grant agreement No 805445.


\bibliography{sample63}{}
\bibliographystyle{aasjournal}



\end{document}